\documentclass[twocolumn,showpacs,preprintnumbers,amsmath,amssymb]{revtex4}

\usepackage{graphicx}
\usepackage{epstopdf}
\usepackage{xspace} 
\usepackage{amsmath}
\usepackage{mathrsfs}
\usepackage{amsfonts}
\usepackage{amssymb}
\usepackage[francais]{babel}
\usepackage[latin1]{inputenc}

\begin{document}

\title{Implication of neutrino backgrounds on the reach of next generation dark matter direct detection experiments}

\author{J.~Billard}\email{billard@mit.edu} \affiliation{Department of Physics, Massachusetts Institute of Technology, Cambridge, MA 02139, USA}
\author{L.~Strigari} \affiliation{Kavli Institute for Particle Astrophysics and Cosmology, Stanford University, Stanford, CA 94305, USA} 
\affiliation{Department of Physics, Indiana University, Bloomington, Indiana 47405, USA}
\author{E. Figueroa-Feliciano} \affiliation{Department of Physics, Massachusetts Institute of Technology, Cambridge, MA 02139, USA}

\begin{abstract}
As direct dark matter experiments continue to increase in size, they will become sensitive to neutrinos from astrophysical sources. For experiments that do not have directional sensitivity, coherent neutrino scattering from several sources represents an important background to understand, as it can almost perfectly mimic an authentic weakly interacting massive particle (WIMP) signal. Here we explore in detail the effect of neutrino backgrounds on the discovery potential of WIMPs over the entire mass range of 500~MeV to 10~TeV. We show that, given the theoretical and measured uncertainties on the neutrino backgrounds, direct detection experiments lose sensitivity to light ($\sim10$ GeV) and heavy ($\sim100$ GeV) WIMPs with a spin-independent cross section below 10$^{-45}$ and 10$^{-49}$~cm$^2$, respectively.

\end{abstract}


\maketitle


\section{Introduction}
\par Direct dark matter detection experiments searching for the presence of weakly interacting massive particles (WIMPs) are rapidly improving in sensitivity~\cite{Aprile:2012nq,Agnese:2013rvf}, now probing important regimes of well-motivated theoretical extensions of the standard model that include dark matter candidates~\cite{Jungman:1995df,Bertone:2004pz,Strigari:2012gn}. As the sensitivity of the experiments continues to improve, it will become increasingly important to quantify their detection prospects in mass and cross section regimes in which backgrounds affect the detection of a WIMP signal. 

\par Upcoming direct dark matter detection experiments will have sensitivity to detect neutrinos from several astrophysical sources, including the Sun, atmosphere, and diffuse supernovae~\cite{Cabrera:1984rr,Monroe:2007xp,Strigari:2009bq,Gutlein:2010tq,Harnik:2012ni}. Though neither coherent  neutrino scattering nor the WIMP-nucleus interaction have conclusively been observed yet, it is of main interest to estimate how the neutrino signal could impact a potential WIMP detection.  For example, the $^8$B solar neutrinos induce an event rate equivalent to a WIMP of 6~GeV/c$^2$ with a spin-independent cross section on the nucleon of $\sim5\times10^{-45}$~cm$^2$, while the atmospheric neutrino background induces an event rate that is similar to a $\sim 100$~GeV WIMP with a spin-independent cross section of $\sim10^{-48}$~cm$^2$~\cite{Strigari:2009bq}. Therefore, if the WIMP mass and cross section are at these scales, neutrino backgrounds must be taken into account when attempting to identify a WIMP signal. As discussed in Sec.~\ref{sec:Discovery}, understanding in detail the energy spectrum of the astrophysical neutrino sources in direct detection experiments is necessary to maximize the discovery potential of upcoming experiments. 

\par In this paper, we systematically quantify the discovery potential of direct dark matter searches in the presence of neutrino backgrounds. In comparison with previous studies, we extend the calculation of the neutrino backgrounds to both lower detectable recoil energies and neutrino fluxes. We consider all realistically feasible configurations of germanium and xenon experiments, and statistically quantify the discovery potential for each experimental configuration as a function of WIMP mass and cross section.   

\par This paper is organized as follows: In Sec.~\ref{sec:NeutrinoFluxes} we discuss and update the neutrino fluxes that we use in our calculations. In Sec.~\ref{sec:Rates} we determine the neutrino-induced recoil event rate and compare to the expected event rate from canonical WIMP models. In Sec.~\ref{sec:Reconstruction} we introduce a new statistical methodology to extract the neutrino background and compare it to an expected WIMP signal. In Sec.~\ref{sec:Discovery} we present the discovery limits as a function of WIMP mass and cross section. Finally, Sec.~\ref{sec:map} discusses the evolution of the discovery potential as a function of exposure and presents the discovery potential of future high exposure direct detection experiments. The last section is dedicated to our conclusions and discusses future directions for direct detection of dark matter.


\section{Neutrino Fluxes}
\label{sec:NeutrinoFluxes}
\par Direct detection experiments will be sensitive to the flux of solar, atmospheric, and diffuse supernova neutrinos. In this section we discuss the respective neutrino fluxes, updating the input from previous calculations, and examining their respective uncertainties.

\subsection{Solar neutrinos}
\par Direct dark matter detection experiments that are sensitive to neutrino-nucleus coherent scattering are primarily sensitive to two sources of solar neutrinos, so called $^8$B and $hep$ neutrinos. The $^8$B neutrinos arise from the decay $^8\textrm{B} \rightarrow {}^7 \textrm{Be}^{*} + e^{+} + \nu_e$, which occurs in approximately 0.02\% of the terminations of the proton-proton ($pp$) chain. The total flux measured with the neutral current (NC) interaction of $^8$B solar neutrinos is $\phi_{NC} = 5.09\pm0.64\times10^6$~cm$^{-2}$~s$^{-1}$ (about 16\% uncertainty)~\cite{Ahmad:2002jz}. Our calculations use the theoretical value $\phi_{NC} = 5.69 \pm0.61\times10^6$ \textrm{cm}$^{-2}$~s$^{-1}$ of the solar neutrino fluxes from Ref.~\cite{Bahcall:2004pz}. This is near the flux prediction of the high metallicity standard solar model (SSM), and thus provides a conservative estimate of the $^8$B neutrino background in dark matter detectors. Note that the low metallicity solution predicts a lower value of the $^8$B  flux normalization, which is statistically inconsistent with the high metallicity SSM (for a detailed discussion see Ref.~\cite{Robertson:2012ib}). The $hep$ neutrinos arise from the reaction $^3\textrm{He} + \textrm{p} \rightarrow ^4\textrm{He} + e^{+}  + \nu_e$, which occurs in approximately $2 \times 10^{-5}$\% of the terminations of the $pp$ chain. At the lowest neutrino energies,  electron capture reaction on $^7$Be is the second largest neutrino source that leads to two monoenergetic neutrino lines at 384.3 and 861.3 keV with a branching ratio of 10\% and 90\% respectively due to the $^7$Li excited state. According to the BS05(OP) solar model, we chose a  $^7$Be  neutrino flux of $4.84\times 10^9$~cm$^{-2}$~s$^{-1}$ with a theoretical uncertainty of about 10.5\% \cite{Bahcall:2004pz}. For the analysis in this paper we are also sensitive to carbon-nitrogen-oxygen cycle (CNO) neutrinos. The uncertainty in the solar composition is the dominant source of uncertainty in the CNO neutrino fluxes. We take an uncertainty of 30\% on the CNO neutrino fluxes~\cite{Serenelli:2011py,PenaGaray:2008qe}.

\par Through neutrino-electron scattering, dark matter detection experiments are also sensitive to neutrinos produced directly in the $pp$ chain. The total flux of neutrinos produced in the $pp$ chain is $5.94 \times 10^{10}$~cm$^{-2}$~s$^{-1}$. Because the neutrino-electron scattering cross section is flavor dependent, in this case we must consider the flavor composition of the neutrino flux that arrives on the Earth. For the energies that we are sensitive to, the electron neutrino survival probability is approximately 55\%~\cite{Lopes:2013nfa}. Following Ref.~\cite{Bahcall:2004pz}, we will consider an uncertainty of 1\% on the  $pp$ neutrino flux.

 \begin{figure}[t]
\begin{center}
\includegraphics[scale=0.44,angle=0]{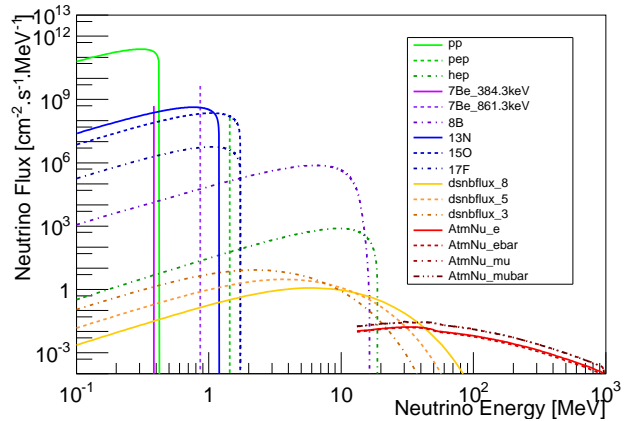}
\caption{Relevant neutrino fluxes which are backgrounds to direct dark matter detection experiments: Solar, atmospheric, and diffuse supernovae~\cite{Strigari:2009bq}.} 
\label{fig:Flux}
\end{center}
\end{figure} 

\begin{center}
 \begin{figure*}[t]
\includegraphics[scale=0.44,angle=0]{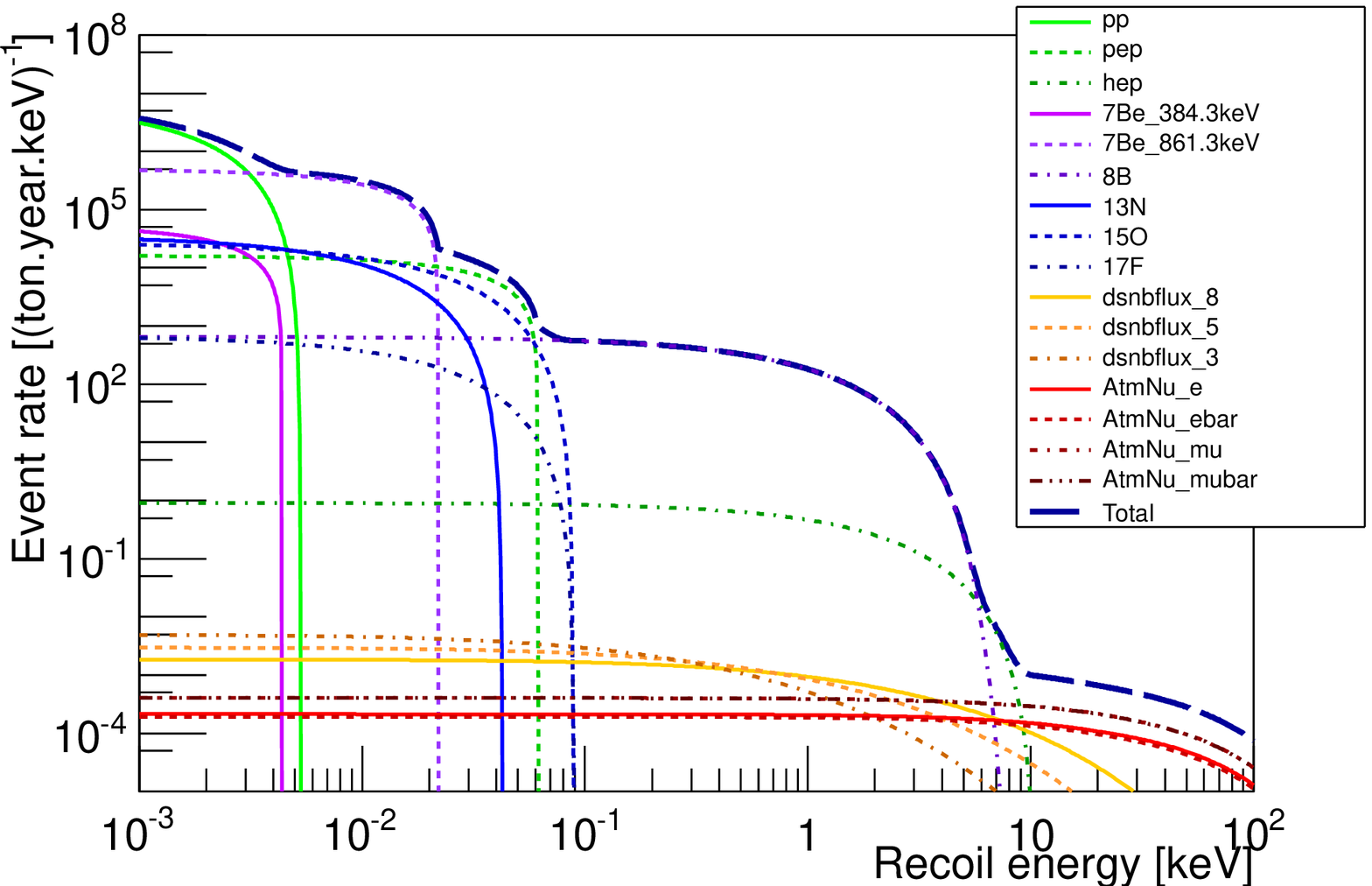}
\includegraphics[scale=0.44,angle=0]{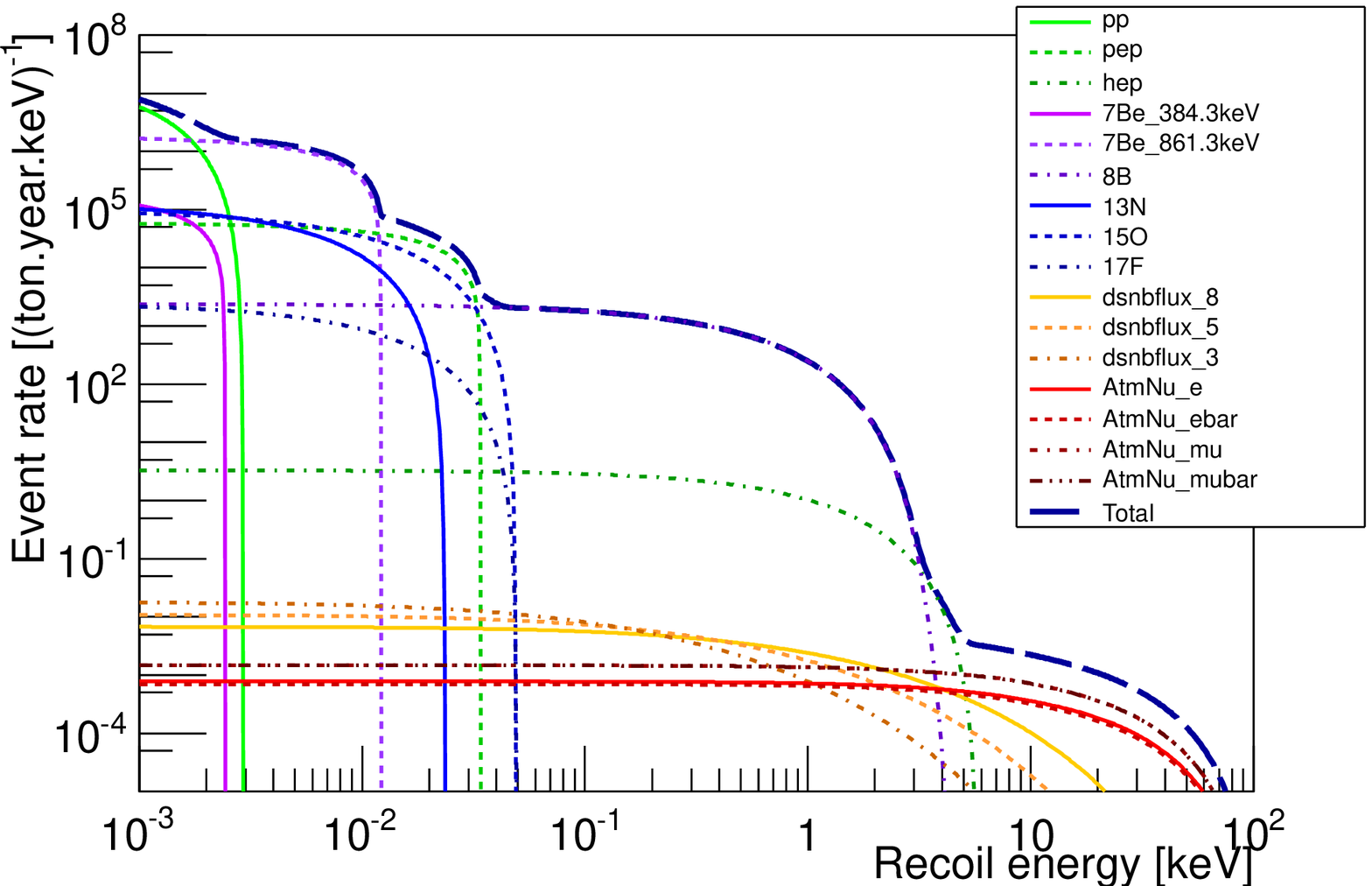}
\caption{Neutrino-induced nuclear recoil spectra for the different neutrino sources, for a Ge target (left) and a Xe target (right).} 
\label{fig:Rates}
\end{figure*} 
\end{center}

\subsection{Atmospheric neutrinos}
\par Atmospheric neutrinos are produced through cosmic ray collisions in the Earth's atmosphere. The collisions produce pions which then decay to muon and electron neutrinos and antineutrinos. The atmospheric neutrino flux has been detected by several experiments: Super-Kamiokande~\cite{Fukuda:1998ub}, SNO~\cite{Aharmim:2009zm}, MINOS~\cite{Adamson:2012gt}, and IceCube~\cite{Abbasi:2010ie}. In these experiments, the direction of the detected muon is reconstructed. Modern direct dark matter detectors do not have directional sensitivity and are mainly sensitive to the low component of the atmospheric neutrino flux, {\it i.e.} less than approximately 100 MeV. At these energies, the uncertainty on the predicted atmospheric neutrino flux is approximately 20\%~\cite{Honda:2011nf}. Due to a cutoff in the rigidity of cosmic rays induced by the Earth's geomagnetic field at low energies, the atmospheric neutrino
flux is larger for detectors that are nearer to the poles~\cite{Honda:2011nf}.

\subsection{Diffuse supernova neutrinos}
\par The diffuse supernova neutrino background (DSNB) is the flux from the past history of all supernova explosions in the Universe. The DSNB flux is a convolution of the core-collapse supernova rate as a function of redshift with the neutrino spectrum per supernova. The core-collapse rate is derived from the star-formation rate and stellar initial mass function; for a recent review on the predicted DSNB flux see Beacom~\cite{Beacom:2010kk}. The neutrino spectrum of a core-collapse supernova is believed to be similar to a Fermi-Dirac spectrum, with temperatures in the range 3-8 MeV. The calculations in this paper assume the following temperatures for each neutrino flavor: T$_{\nu_e}$ = 3 MeV, T$_{\bar \nu_e}$ = 5 MeV, and T$_{\nu_x}$ = 8 MeV. Here T$_{\nu_x}$ represent the remaining four flavors: $\nu_\mu$, $\bar{\nu}_\mu$, $\nu_\tau$, and $\bar{\nu}_\tau$. Because of the scaling of the coherent neutrino scattering cross section (integrated over all recoil energies), the flavors with the largest temperature dominate the event rate. Following \cite{Beacom:2010kk}, we will consider a systematic uncertainty on the DSNB flux of 50\%.

\par Figure~\ref{fig:Flux} presents the relevant neutrino fluxes that will be a background for dark matter direct detection. Shown are the different contributions from solar, atmospheric, and diffuse supernova neutrinos. Note that we are not considering geoneutrinos nor reactor neutrinos in this study. Indeed, as shown in~\cite{Monroe:2007xp}, the contribution of the geoneutrinos to the neutrino-induced recoil energy spectrum is at least 2 orders of magnitude below the solar neutrino contribution over the whole energy range. The reactor neutrinos are strongly dependent on the location of the experiment with respect to the surrounding nuclear reactors and on the power these reactors are working at. While this contribution should be estimated independently for each experiment, we are not considering them as this is beyond the scope of this paper and will therefore only discuss the case of cosmic neutrinos as shown in Fig.~\ref{fig:Flux}.


\section{WIMP and neutrino background event rate calculations}
\label{sec:Rates}

\subsection{WIMP-induced nuclear recoil rate calculation}

Like most spiral galaxies, the Milky Way is believed to be immersed in a halo of WIMPs which outweighs the luminous component by at least an order of magnitude \cite{salucci, klypin,Strigari:2012gn}. The velocity distribution of dark matter in the halo is traditionally modeled as a Maxwell-Boltzmann, characterized by a density profile that scales as $1/r^{2}$ and leading to the observed flat rotation curve~\cite{lewin}. Recent results from N-body simulations in fact indicate that this Maxwell-Boltzmann assumption is an oversimplification~\cite{Vogelsberger:2008qb,Kuhlen:2009vh,Mao:2012hf}, as there is a wider peak and there are fewer particles in the tail of the distribution; this result has important implications for interpretation of experimental results~\cite{Mao:2013nda}. Further, substructures, streams, and a dark disk may create distinct features in the velocity distribution~\cite{Nezri, Ling, Bruch, Read}. Since the goal of this paper is to examine the effects of the neutrino background on the extraction of a WIMP signal, to make the connection to previous experimental studies in this paper we just consider the Maxwell-Boltzmann model, which is characterized by the following WIMP velocity distribution in the Earth frame,
\begin{equation}
f(\vec{v}) = \left\{
\begin{array}{rrll}
\rm & \frac{1}{N_{\rm esc}(2\pi\sigma^2_v)^{3/2}}\exp\left[-\frac{\left(\vec{v} + \vec{V}_{\rm lab}\right)^2}{2\sigma^2_v}\right] &	\ \text{if $|\vec{v} + \vec{V}_{\rm lab}|<v_{\rm esc}$}  \\
\rm & 0  		& 	\ \text{if $|\vec{v} + \vec{V}_{\rm lab}|\geq v_{\rm esc}$}
\end{array}\right.
\end{equation}
where $\sigma_v$ is the WIMP velocity dispersion related to the local circular velocity $v_0$ such that $\sigma_v = v_0/\sqrt{2}$, $\vec{V}_{\rm lab}$ and $v_{\rm esc}$ are respectively the laboratory and the escape velocities with respect to the galactic rest frame, and $N_{\rm esc}$ is the correction to the normalization of the velocity distribution due to the velocity cutoff ($v_{\rm esc}$). \\

The differential recoil energy rate is then given by \cite{lewin},
\begin{equation}
\frac{dR}{dE_r} = M T\times \frac{\rho_0 \sigma_0}{2m_{\chi}m^2_r}F^2(E_r)\int_{v_{\rm min}}\frac{f(\vec{v})}{v}d^3v
\end{equation} 
where $\rho_0$ is the local dark matter density, $m_{\chi}$ is the WIMP mass, $m_r = m_\chi m_N/(m_\chi + m_N)$ is the WIMP-nucleus reduced mass and $\sigma_0$ is the normalized to nucleus cross section. Note that we will assume that the WIMP couples identically to the neutrons and protons, though generically a larger theoretical parameter space is available~\cite{Feng:2011vu}. $F(E_r)$ is the nuclear form factor that describes the loss of coherence for recoil energies above $\sim$10~keV. In the following, we will consider the standard Helm form factor \cite{lewin}. For the sake of comparison with running experiments, we will consider the standard values of the different astrophysical parameters: $\rho_0 = 0.3 $~GeV/c$^2$/cm$^3$, $v_0 = 220$~km/s, $V_{\rm lab} = 232$~km/s and $v_{\rm esc} = 544$~km/s.

\begin{center}
 \begin{figure*}[t]
\includegraphics[scale=0.44,angle=0]{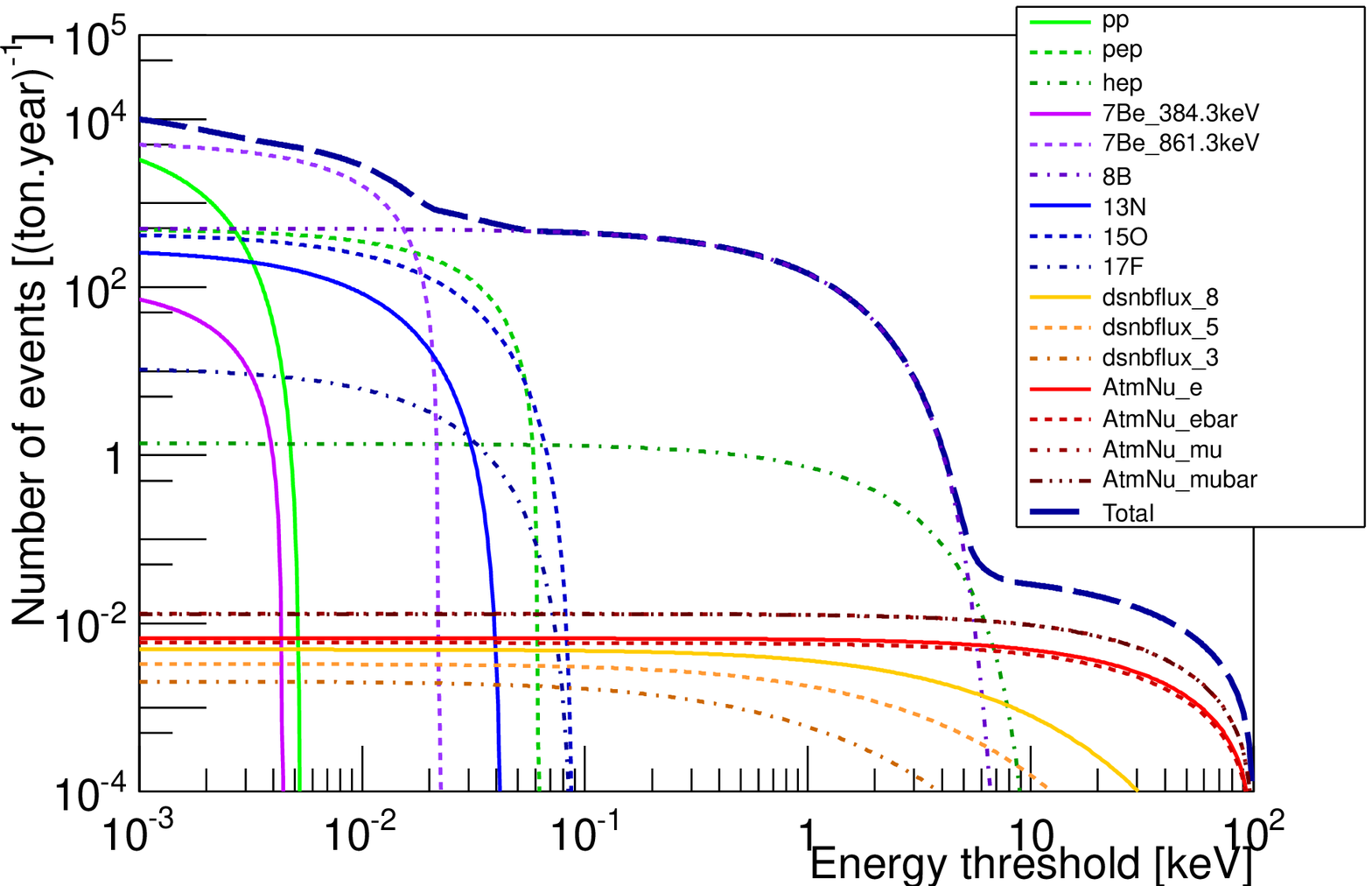}
\includegraphics[scale=0.44,angle=0]{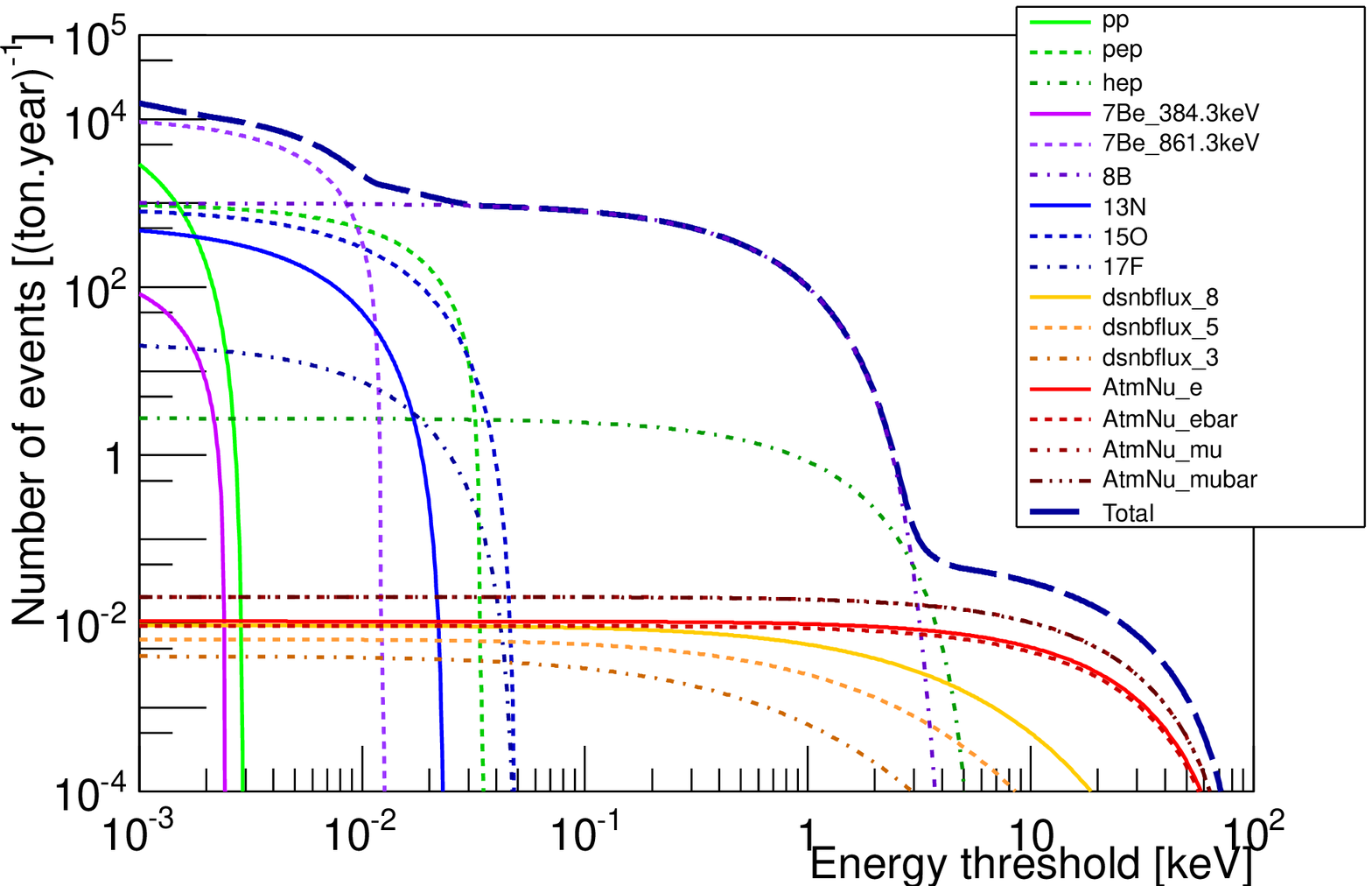}
\caption{Number of neutrino-induced nuclear recoils per ton-year for a Ge target (left) and Xe target (right) as a function of the energy threshold. Note that we have considered an upper limit on the nuclear recoil energy range of 100 keV.} 
\label{fig:Number}
\end{figure*} 
\end{center}

\subsection{Neutrino-nucleus cross section}
It has been shown by Freedman \cite{freedman} that the neutrino-nucleon elastic interaction, well explained by the standard model, leads to a coherence effect implying a neutrino-nucleus cross section that approximately scales as the atomic number ($A$) squared. However, this coherent nature of the neutrino-nucleus scattering can only take place when the momentum transfer is comparable in scale to the nuclear size, about a few keV for most targets of interest. At higher energies, generally above 10 keV for most nuclei, the loss of coherence reduces the neutrino-nucleus cross section. The resulting differential neutrino-nucleus cross section as a function of the recoil energy $E_r$ and the neutrino energy $E_\nu$ is defined as follows \cite{freedman2}:
\begin{equation}
\frac{d\sigma(E_\nu, E_r)}{dE_r} = \frac{G^2_f}{4\pi}Q^2_w m_N \left(1 - \frac{m_NE_r}{2E^2_{\nu}}  \right)F^2(E_r),
\end{equation}
where $m_N$ is the target nucleus mass, $G_f$ is the Fermi coupling constant, $Q_w = N - (1-4\sin^2\theta_w)Z$ is the weak nuclear hypercharge with $N$ the number of neutrons, $Z$ the number of protons, and $\theta_w$ the weak mixing angle. \\
From kinematics, one can easily derive that the maximum recoil energy $E^{\rm max}_r$ is equal to:
\begin{equation}
 E^{\rm max}_r = \frac{2E^2_\nu}{m_N+2E_\nu}. 
\label{eq:maxE}
\end{equation}
It is worth noticing that for neutrino energies above $\sim$100~MeV some additional processes start to contribute to the total cross section, such as quasielastic scattering, resonance production, and deep inelastic scattering for higher energies \cite{Formaggio:2013kya}. However, as their contribution to the event rate in the recoil energy range of interest is negligible, we will neglect these additional contributions in the following of this study.

\subsection{Neutrino-electron cross sections}
Neutrino-induced electronic recoils can be an important background for upcoming ton-scale experiments that do not reach sufficiently high power in electronic recoil background rejection. In this case, such background processes should be accounted for in the estimation of the discovery reach of these experiments. In the following, we will discuss the two main neutrino-electron scattering processes that are relevant for neutrino energies below 1-10 MeV~\cite{Marciano:2003eq}, the standard electroweak interaction and the neutrino magnetic moment. \\

As the $pp$ neutrinos provide the dominant contribution to the solar neutrino flux and the maximum recoil energy induced by these neutrinos is about 260~keV, we can safely neglect the other neutrino components to the total neutrino-induced electronic recoil background.  Also, in the following calculations, we will neglect atomic effects and consider the electrons from the atomic cloud as being free \cite{Marciano:2003eq,henry}. 

 \begin{figure*}[t]
\begin{center}
\includegraphics[scale=0.44,angle=0]{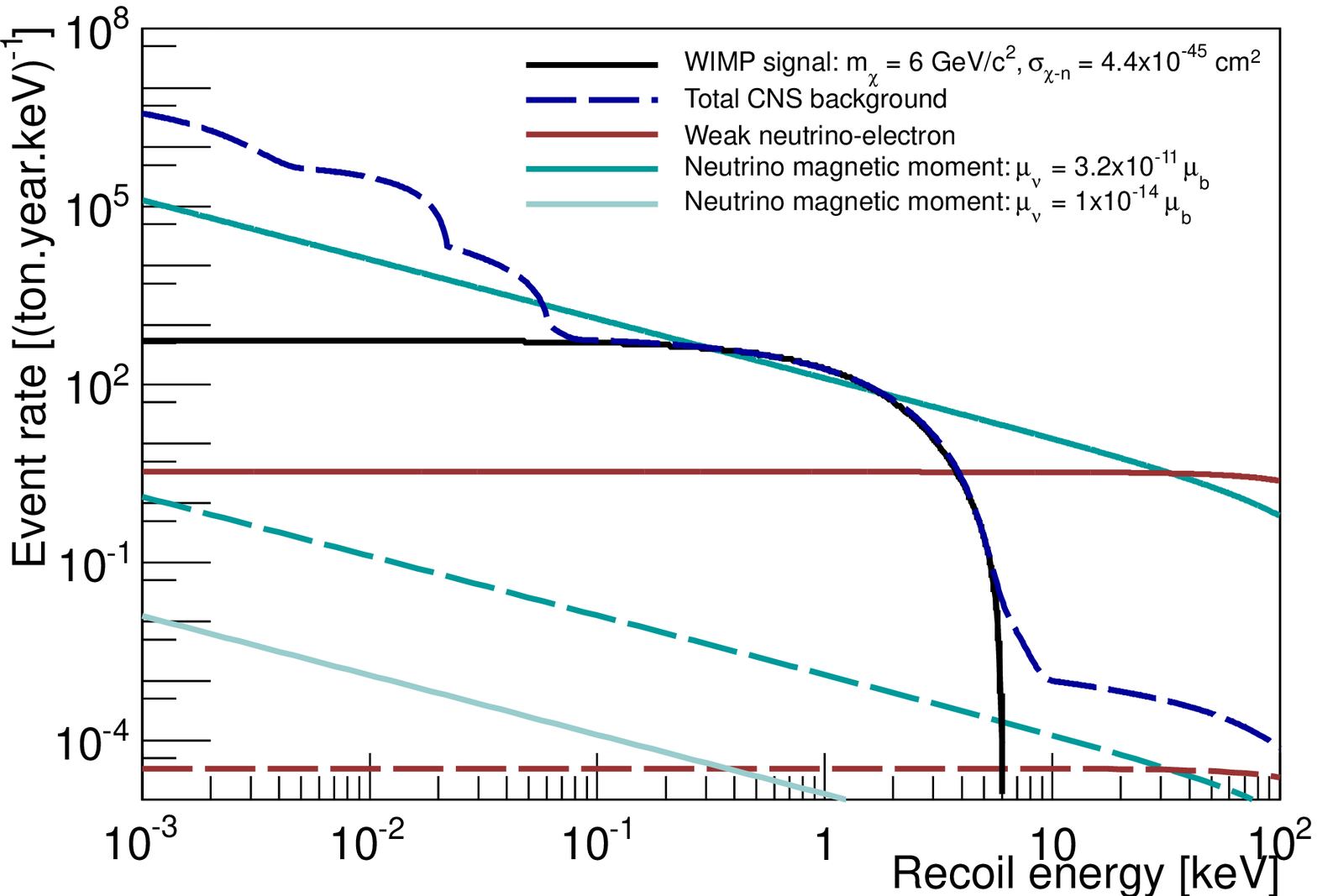}
\includegraphics[scale=0.44,angle=0]{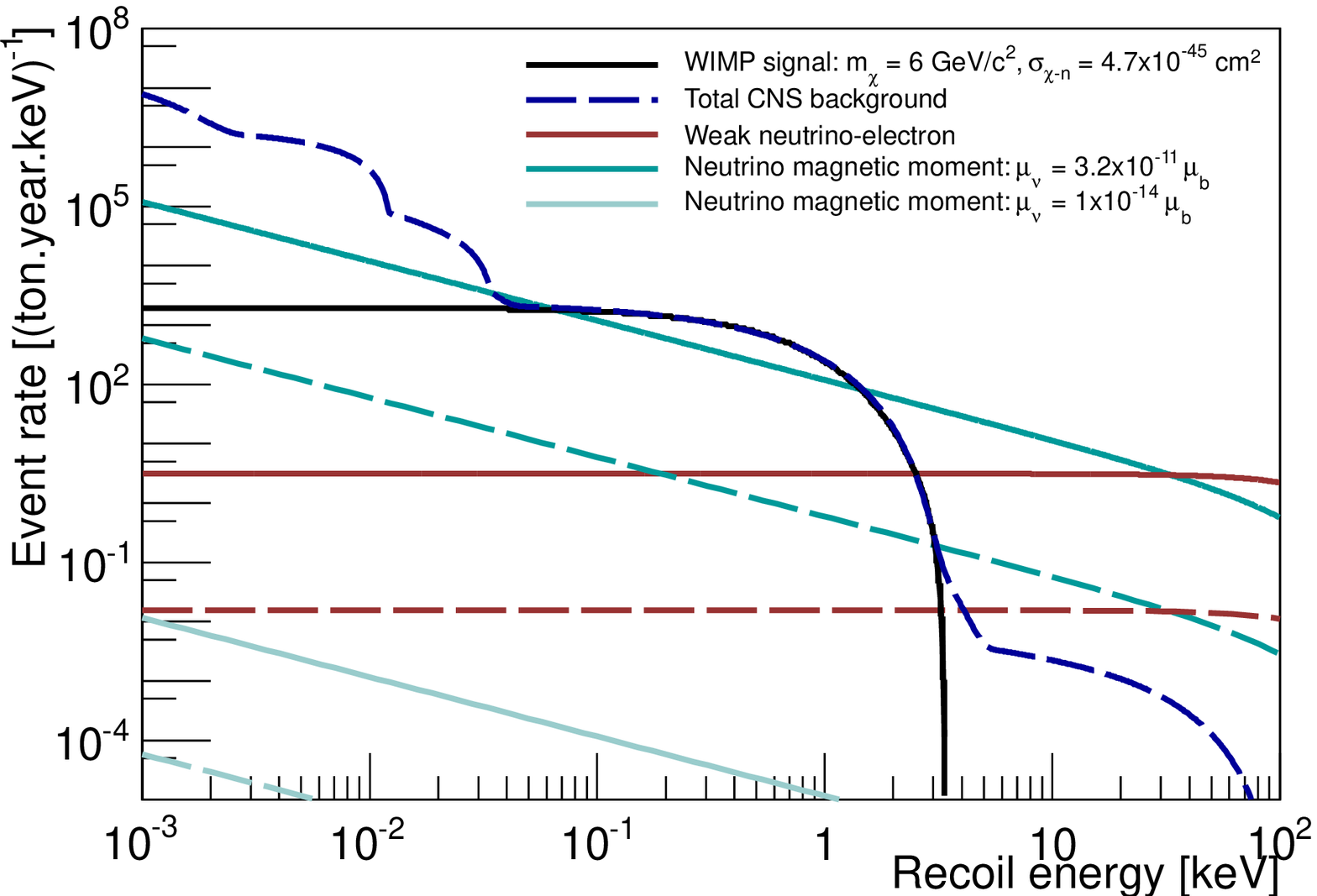}
\caption{Left (right) panel shows the energy spectra of the most relevant neutrino backgrounds for a Ge (Xe) type detector. Shown are a WIMP signal (black solid line), the total coherent neutrino scattering (CNS) background contribution (blue dashed line), standard electroweak neutrino-electron interaction (red line) and the contribution from the neutrino magnetic moment (cyan lines). Dashed red and cyan lines (dark and light) correspond to the consideration of an electron rejection factor of 99.5\% and 10$^5$ for a XENON-like and Ge-based CDMS-like experiment respectively. Dark and light cyan curves correspond to the experimental and theoretical upper limits on the neutrino magnetic moment respectively.} 
\label{fig:ERback}
\end{center}
\end{figure*} 

\subsubsection{Standard electroweak interaction}

At tree level, the neutrino-electron electroweak interaction proceeds through the exchange of a $Z$ boson (neutral current) and the exchange of a $W$ boson (charged current) which is only possible in the case of an incoming electron neutrino. The resulting expression of the cross section is as follows \cite{Marciano:2003eq,Formaggio:2013kya}:
\begin{align}
\frac{d\sigma(E_\nu, E_r)}{dE_r} = & \frac{G_f^2m_e}{2\pi}\left[ (g_v + g_a)^2 \right. \nonumber \\ 
&\left. + (g_v - g_a)^2\left(1 - \frac{E_r}{E_\nu}\right)^2 + (g_a^2 - g_v^2)\frac{m_e E_r}{E_\nu^2}  \right], 
\end{align}
where $m_e$ is the electron mass, $g_v$ and $g_a$ are the vectorial and axial coupling respectively and are defined such that:
\begin{equation}
g_v = 2\sin^2\theta_w - \frac{1}{2} \ \ \ \ \ \ \ g_a = \frac{1}{2}. 
\end{equation}
In the particular case $\nu_e + e \rightarrow \nu_e + e$, the interference due to the additional charged current  contribution implies a shift in the vectorial and axial coupling constants such that $g_{v,a} \rightarrow g_{v,a} + 1$. One can easily derive that the $\nu_e + e \rightarrow \nu_e + e$ cross section is about one order of magnitude larger than in the case of $\nu_l + e \rightarrow \nu_l + e$ (where $l = \mu, \tau$). Hence, it is important to consider the neutrino oscillation from the solar core to the Earth-based detector when computing this neutrino-electron background. It has been shown in Ref.~\cite{Lopes:2013nfa} that the survival probability of $\nu_e$ below 1 MeV is fairly constant in energy and equal to 0.55. The remaining component is distributed between $\nu_{\mu}$ and $\nu_{\tau}$ which have the same expression of the cross section.

\subsubsection{Neutrino magnetic moment}

As neutrinos oscillate, they must have a non-vanishing mass and sufficiently large mixing with each other. In the case of a Dirac neutrino, the extension of the standard model in which neutrinos are massive naturally provides a small but nonzero neutrino magnetic moment. This results in an increase of the total neutrino-electron scattering cross section by the following contribution~\cite{Marciano:2003eq,vogel}:
\begin{equation}
\frac{d\sigma(E_\nu, E_r)}{dE_r} = \mu_\nu^2\frac{\pi \alpha^2}{m_e^2}\left[ \frac{1}{E_r} - \frac{1}{E_\nu} \right], 
\end{equation}
where $\mu_\nu$ is the neutrino magnetic moment in units of Bohr magneton $\mu_b = e/2m_e$ and $\alpha$ is the fine structure constant. The simplest standard model prediction leads to a very tiny magnetic moment of about $\mu_\nu \sim 10^{-20} \mu_b$ preventing any experiment from being sensitive to this putative contribution. However, some more general extensions could predict neutrino magnetic moment up to about $\mu_\nu \sim 10^{-14}\mu_b$ where Majorana neutrinos generally have higher magnetic moment than Dirac neutrinos \cite{bell,bell2}. As the measurement of such process could therefore be an excellent probe for new physics beyond the standard model, it is of great interest trying to measure it. The strongest experimental upper limit on the neutrino magnetic moment coming from the GEMMA Collaboration is equal to $3.2\times10^{-11}\mu_b$ \cite{Beda:2010hk} ($5\times10^{-12}\mu_b$~\cite{henry}) without (with) considering atomic effects. Evidence of $\mu_\nu > 10^{-14}\mu_b$ would strongly be in favor of new physics at the TeV scale or beyond and would imply that the neutrino is Majorana~\cite{bell2}.

\subsection{Neutrino-induced background rate calculation}
\label{sec:nurate}

 \begin{figure*}[t]
\begin{center}
\includegraphics[scale=0.44,angle=0]{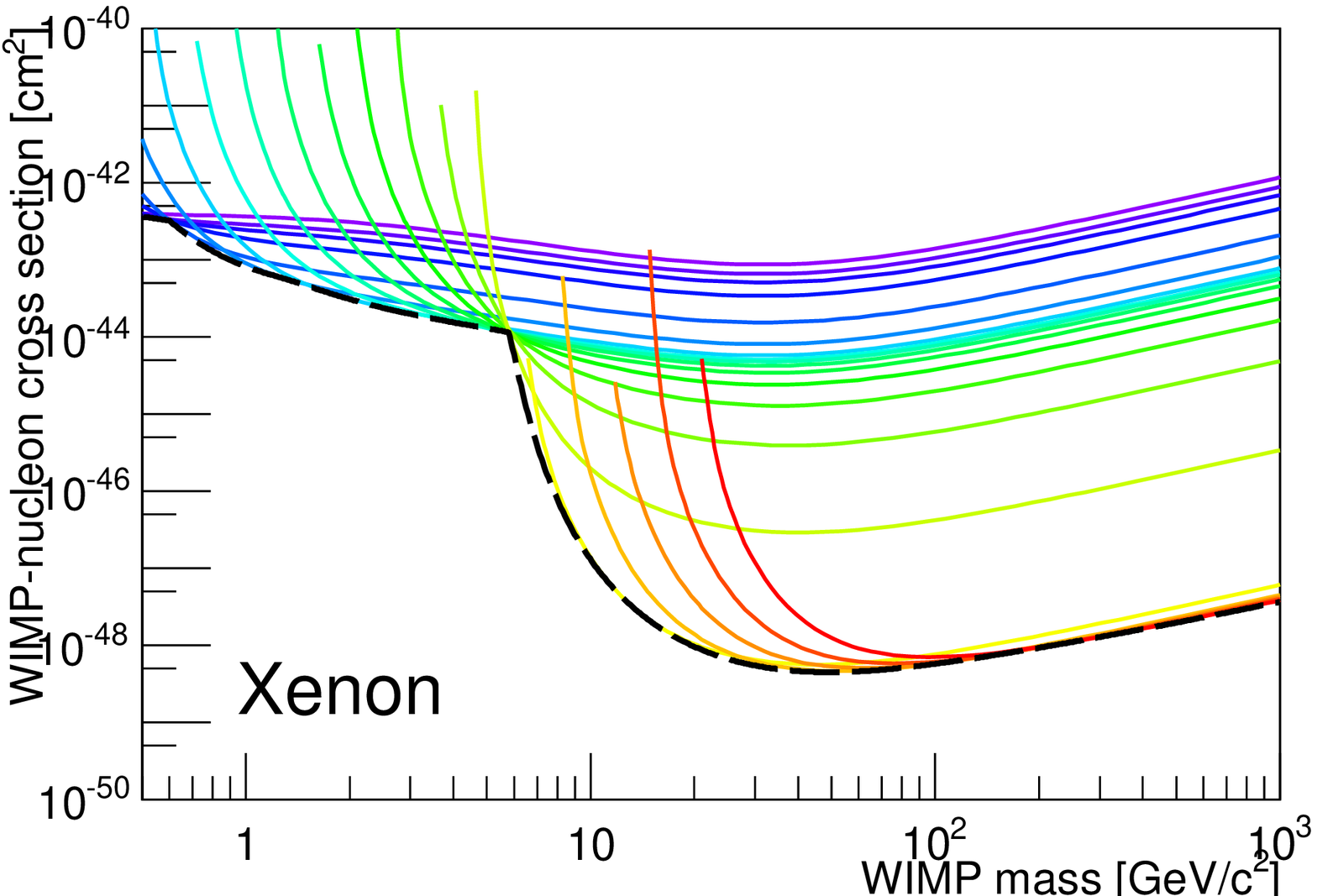}
\includegraphics[scale=0.44,angle=0]{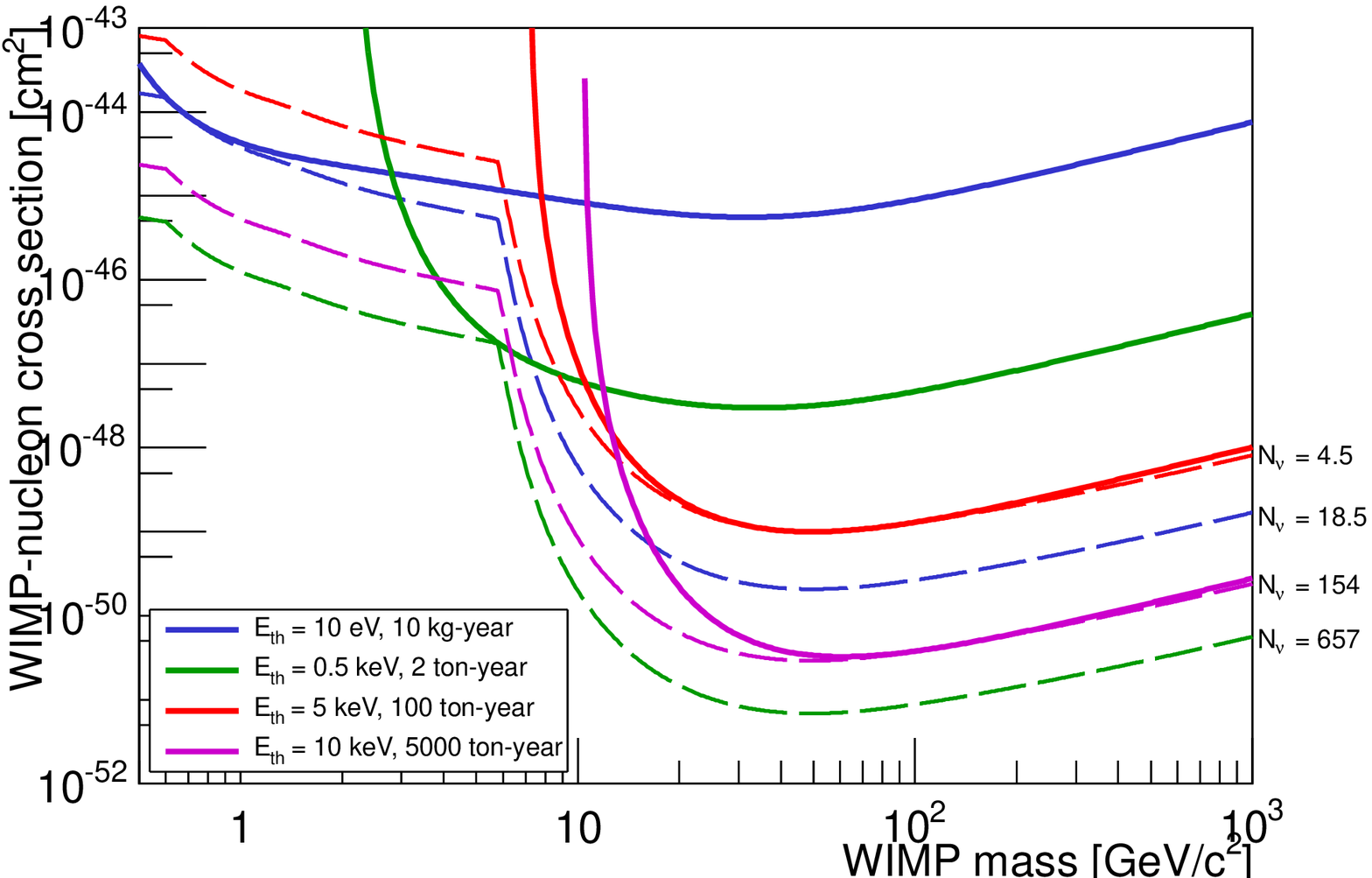}
\caption{Left: Set of derived background-free sensitivity curves for exposures that attain one neutrino event, for different thresholds from 0.001 (purple) to 100~keV (red) in logarithmic steps. The black line is constructed by joining the best sensitivity for each mass, and represents a one neutrino event contour line in the WIMP-nucleon cross section vs WIMP mass plane. Right: Background-free exclusion limits (solid lines) for four different Xe-based experiments with threshold of 10~eV, 500~eV, 5~keV, and 10~keV and exposures of 10~kg-years, 2~ton-years, 100~ton-years, and 5,000~ton-years respectively. Also shown in dashed lines are the neutrino iso-event contour lines for 18.5 (blue), 657 (green), 4.5 (red), and 154 (magenta) events.} 
\label{fig:1neutrinoLine}
\end{center}
\end{figure*}

The calculation of the event rate as a function of the recoil energy is given by:
\begin{equation}
\frac{dR}{dE_r} = \mathscr{N}\times\int_{E^{\rm min}_\nu} \frac{dN}{dE_\nu}\times \frac{d\sigma(E_\nu, E_r)}{dE_r} dE_\nu
\end{equation}
where $dN/dE_\nu$ denotes the neutrino flux and $\mathscr{N}$ is the number of target nuclei per unit of mass of detector material. In the following, we will denote $M$ and $T$ as being respectively the detector mass and the exposure time of the experiment. Note that in the case of the neutrino-induced electronic recoils, the event rate is multiplied by the number of electrons $Z$ per atom. In the limit where $m_N\gg E_\nu$, one can deduce that the minimum neutrino energy $E^{\rm min}_\nu$ required to generate a nuclear recoil at an energy $E_r$ is:
\begin{equation}
 E^{\rm min}_\nu = \sqrt{\frac{m_NE_r}{2}}. 
\end{equation}
However, in the case of an electronic recoil, the expression of $E^{\rm min}_\nu$ is the following:
\begin{equation}
 E^{\rm min}_\nu = \frac{1}{2}\left(E_r + \sqrt{E_r(E_r+2m_e)}  \right). 
\end{equation}

Figure~\ref{fig:Rates} presents the nuclear recoil rate as a function of recoil energy for all neutrino components for a Ge target (left panel) and Xe target (right panel). As shown in Fig.~\ref{fig:Rates}, most of the solar neutrinos are at very low recoiling energies (below 0.1 keV) except the $^8$B and $hep$ neutrinos that will dominate the event rate from 0.1 to 8~keV. Above these energies, atmospheric neutrinos will dominate with a subdominant contribution from the diffuse supernova background neutrinos.

Figure~\ref{fig:Number} presents the expected number of nuclear recoils as a function of the threshold energy and with an upper bound on the recoil energy range of 100 keV.
It is interesting to notice  that the $^8$B neutrinos dominate the expected number of neutrino-induced nuclear recoils for threshold energies between 10 eV and 10 keV. As shown on Fig.~\ref{fig:Number}, a ton-scale experiment with a 0.1~keV threshold can then expect about 500 and 1000 neutrino-induced nuclear recoils for a Ge and Xe based experiment, respectively.

Finally, Fig.~\ref{fig:ERback} presents the total neutrino backgrounds as well as a WIMP spectrum for a benchmark model that best fits the $^8$B neutrino-induced nuclear recoil spectrum (black solid line). It is also interesting to see that a WIMP signal could almost perfectly be mimicked by solar neutrino backgrounds. 
The neutrino background from coherent neutrino scattering is given by the blue dashed line, and the electroweak and neutrino magnetic moment $\nu + e^{-} \rightarrow \nu + e^{-}$ contributions are shown by the solid red and cyan lines. The dark cyan line corresponds to the expected event rate considering the experimental constraint on the neutrino magnetic moment ($\mu_\nu = 3.2\times 10^{-11} \mu_b$) while the light cyan line considers the theoretical upper bound from the most general extensions of the standard model ($\mu_\nu = 10^{-14} \mu_b$). As dark matter experiments aim at rejecting electronic recoils, the dashed red and cyan lines correspond to the event rate expected in a XENON-like experiment where the rejection factor is taken to be flat in energy and equal to 99.5\% \cite{Baudis:2012ig} and equal to $10^5$ in a Ge-based CDMS-like experiment \cite{agnese}. Therefore, after electron recoil rejection, one can easily deduce that neutrino-electron backgrounds should not be an issue for Ge-based CDMS-like experiments while they could contribute significantly to the total neutrino backgrounds for XENON-like experiments for recoil energies above 4~keV. That being said, unless otherwise stated we will only consider neutrino backgrounds from coherent neutrino scattering. \\

For a particular experiment, Fig.~\ref{fig:Number} gives the number of neutrino events for an experiment with a fixed threshold and exposure and a 100\% efficiency over the whole recoil energy range (from the threshold to 100 keV). In this paragraph we present a novel way to represent the level of the neutrino CNS background on the WIMP-nucleon cross section vs WIMP mass plane which is presented for the case of a Xe-based experiment in Fig.~\ref{fig:1neutrinoLine}. To do so, we generated a set of 1,000 background-free exclusion limits, which are defined as isovalues of WIMP events (2.3 at 90\% C.L.), as a function of the WIMP mass, with varying thresholds from 0.001 to 100~keV and adjusted each curve's exposure such that each experiment expects a neutrino background of one event; see colored solid lines in Fig.~\ref{fig:1neutrinoLine} (left).

By taking the lowest cross section from all the limits as a function of the WIMP mass, one can draw the line in the WIMP-nucleon cross section vs WIMP mass plane that corresponds to the best background-free sensitivity estimate achievable at each WIMP mass for a one neutrino event exposure, see black dashed line in Fig.~\ref{fig:1neutrinoLine} (left). This follows from the construction of the line, which joins the mass-dependent threshold/exposure pairs that optimize the background-free sensitivity estimate at each mass while having a background of one neutrino event.

Since both the neutrino background and the background-free WIMP sensitivity scale linearly with exposure (for the same fixed threshold), one can derive the number of expected neutrino events from a given experiment sensitivity by scaling  the one neutrino event line such that it overlaps with the background-free sensitivity limit only at a single point. This is illustrated in Fig.~\ref{fig:1neutrinoLine} (right) where we have shown the background-free sensitivity limits for different Xe-based experiment (solid lines) and their corresponding neutrino isoevent contour lines (dashed lines). The considered thresholds are 10~eV, 500~eV, 5~keV, and 10~keV with exposures of 10~kg-years, 2~ton-years, 100~ton-years and 5,000~ton-years. As one can see from Fig.~\ref{fig:1neutrinoLine} (right), these  experiments expect 18.5, 657, 4.5, and 154 neutrino events respectively. Equivalently, the number of expected neutrino events can be deduced from the maximum ratio over the WIMP mass range between the one neutrino event line and the experiments' background-free limits. A set of contours for Xe-based experiments with a flat efficiency between the threshold up to 100 keV are shown in Fig.~\ref{fig:WIMPlimit} (left). 

 \begin{figure*}[t]
\begin{center}
\includegraphics[scale=0.44,angle=0]{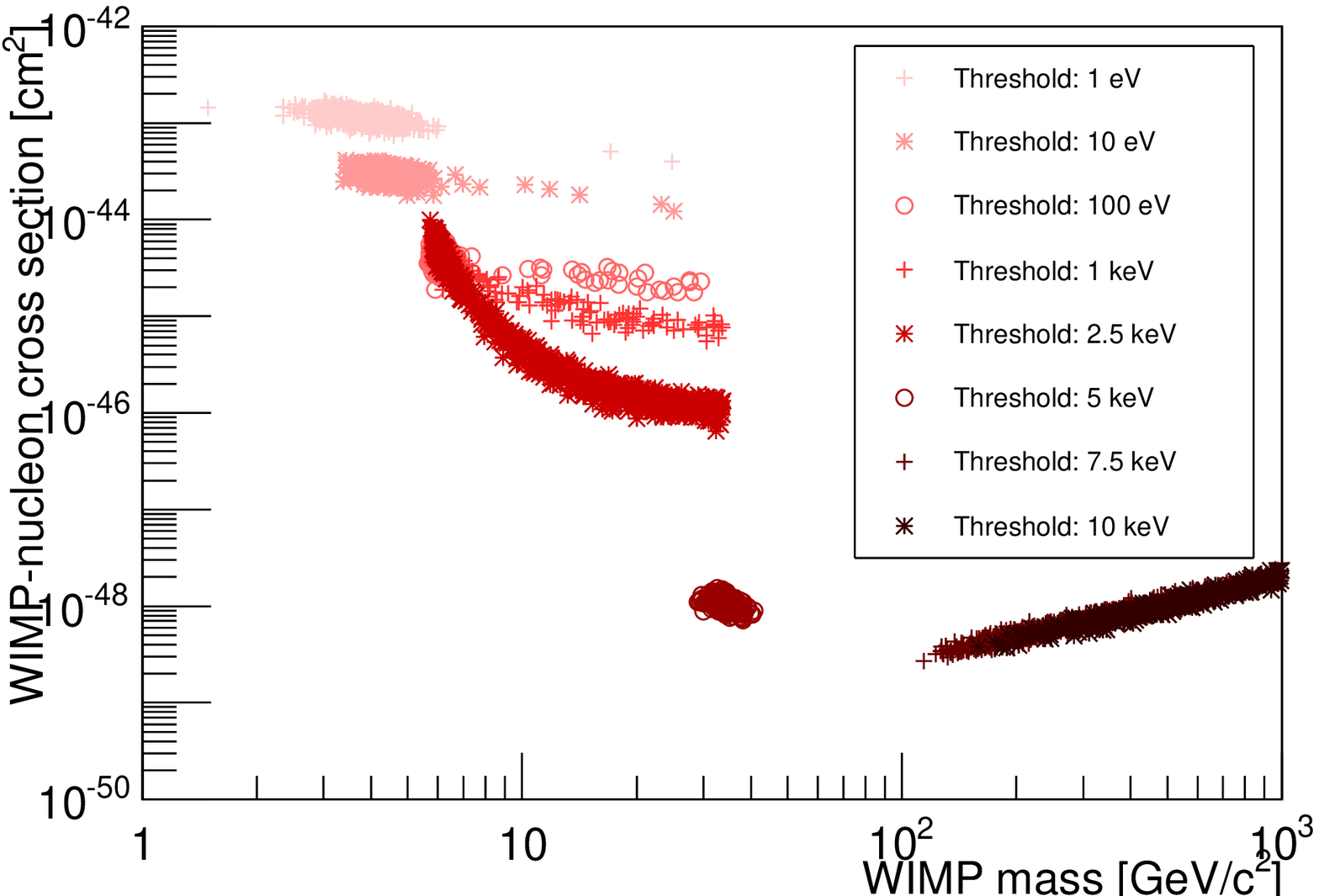}
\includegraphics[scale=0.44,angle=0]{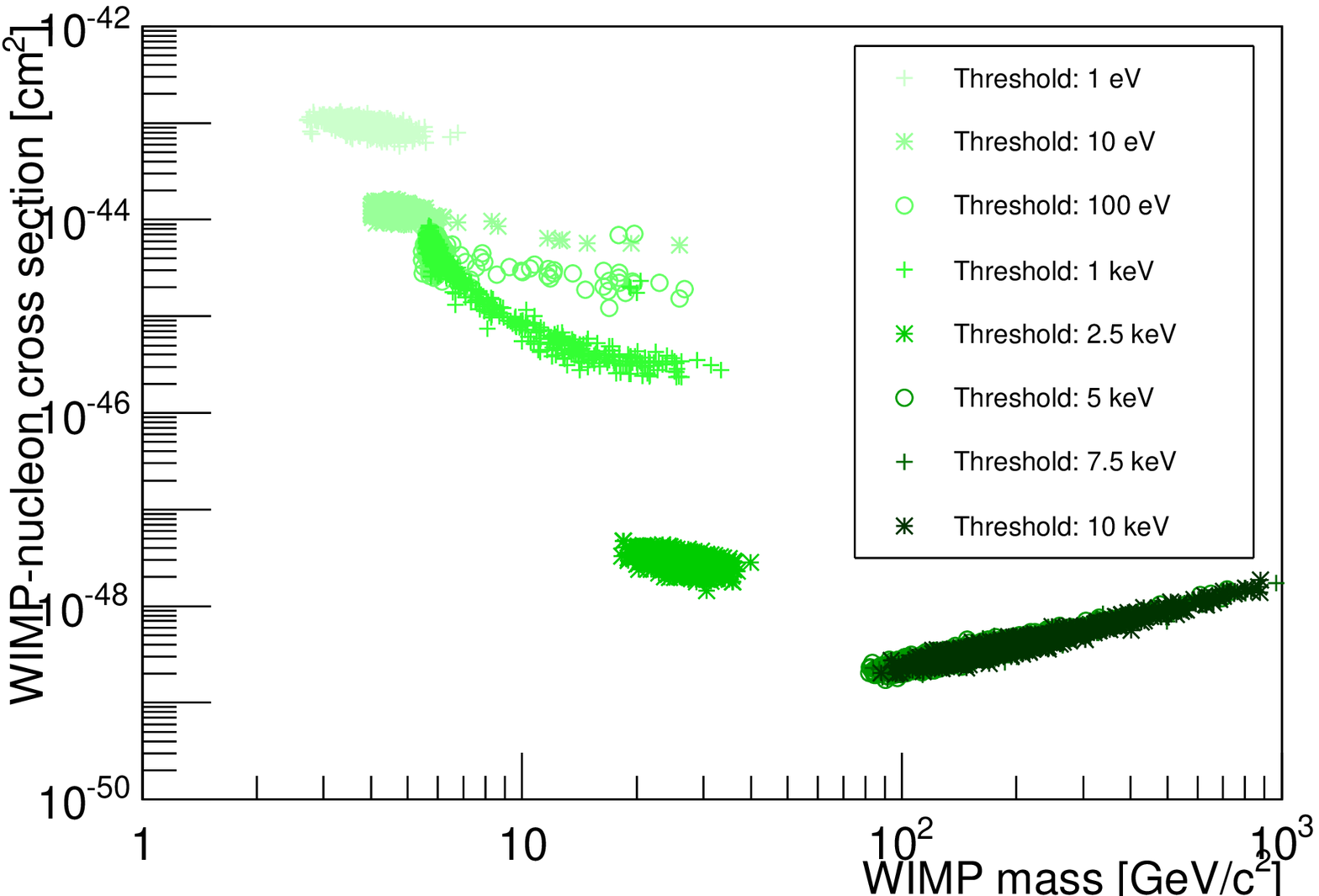}
\caption{Distributions of the maximum likelihood of the CNS background under the WIMP only hypothesis for a Ge target (left) and a Xe target (right). The different intensities of colors correspond to the energy threshold considered, from light to dark: 1 eV, 10 eV, 100 eV, 1 keV, 2.5 keV, 5 keV, 7.5 keV, and 10 keV. These distributions have been computed by adjusting the experiment exposure such that we have a total of about 500 expected neutrino events for each different energy threshold and target.} 
\label{fig:MaxLike}
\end{center}
\end{figure*} 
 \begin{figure*}
\begin{center}
\includegraphics[scale=0.44,angle=0]{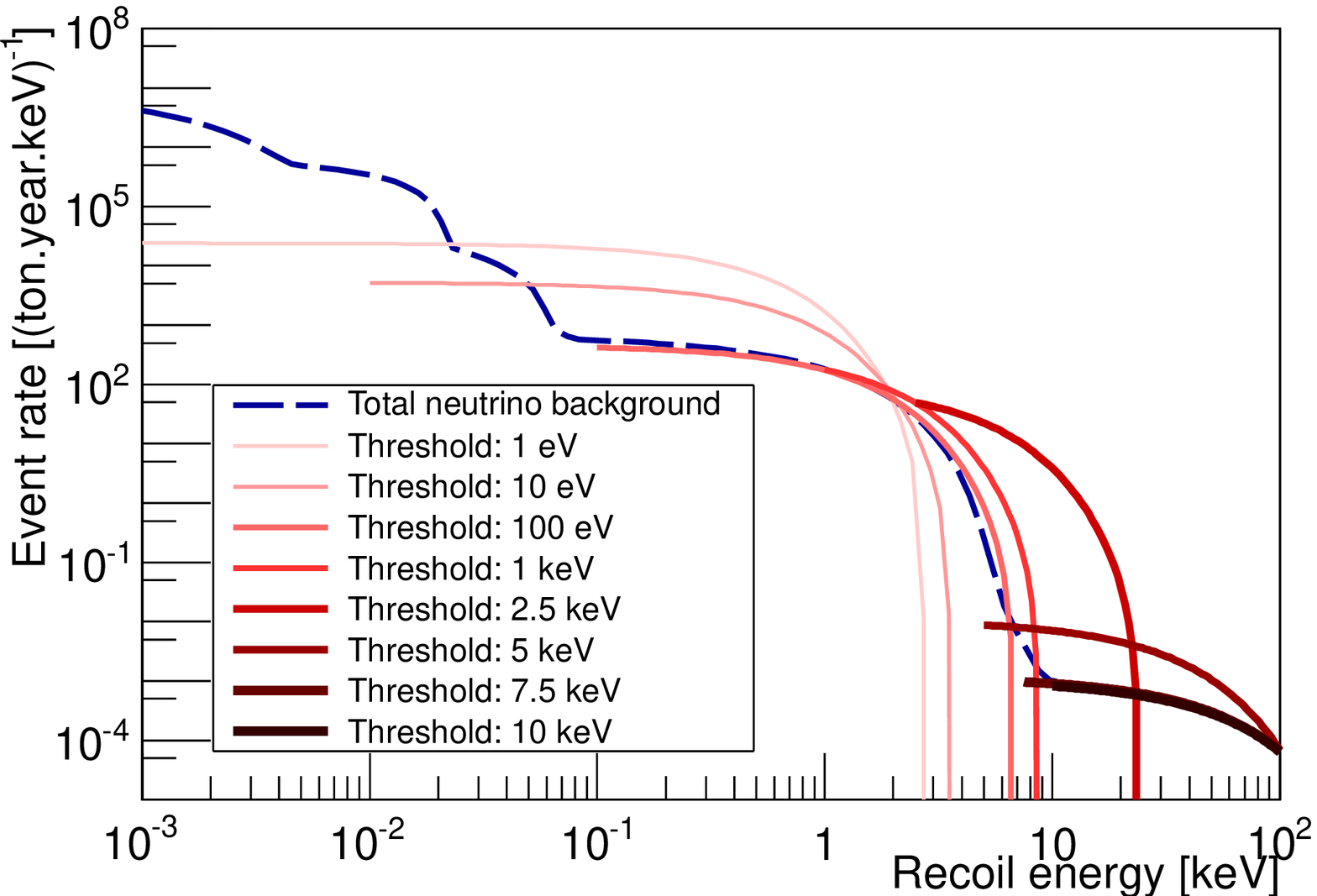}
\includegraphics[scale=0.44,angle=0]{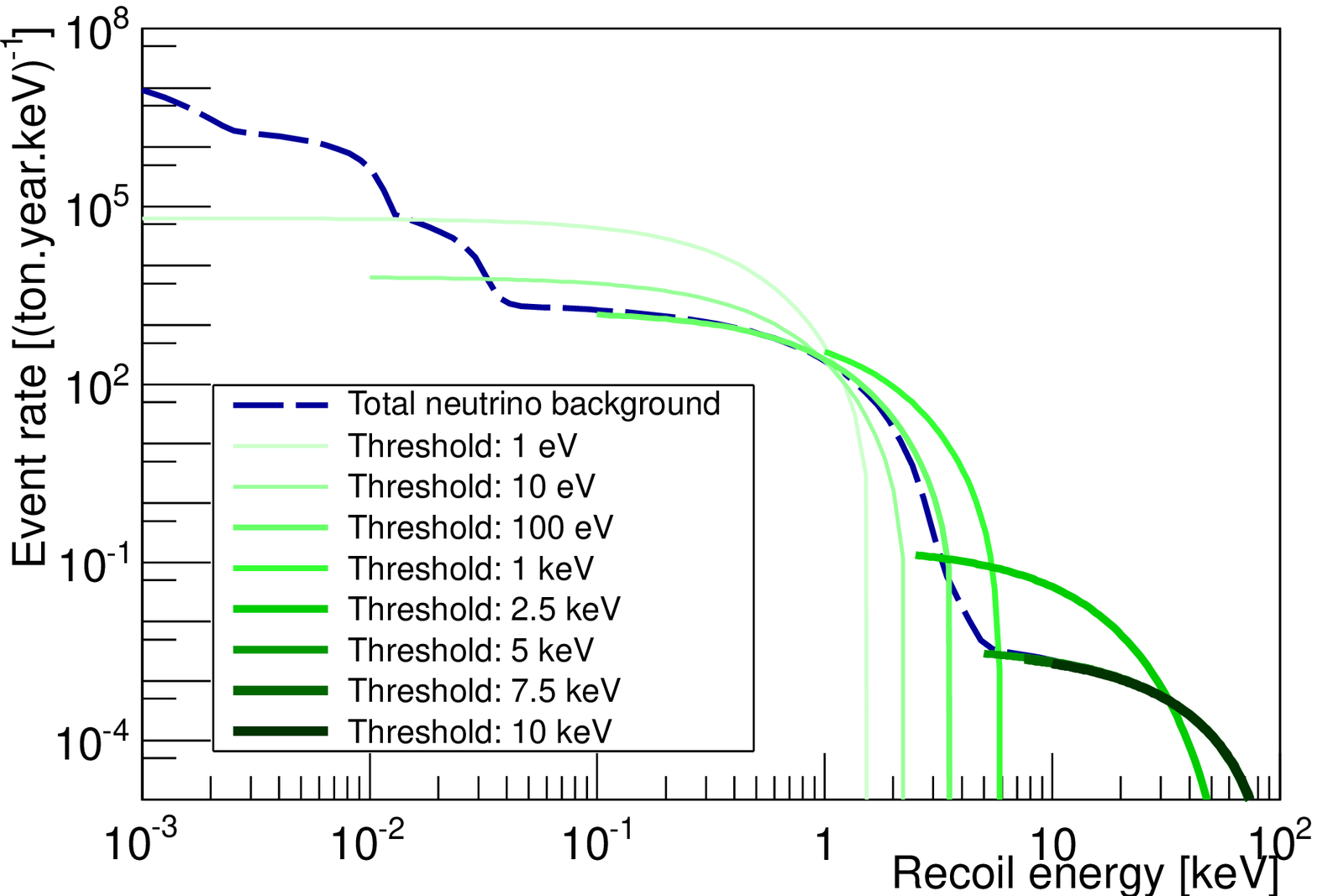}
\caption{Comparison between the nuclear recoil event rate as a function of energy from the CNS background (blue dashed line) and the best fit WIMP models for the different thresholds, deduced from Fig.~\ref{fig:MaxLike}, in the case of a Ge target (left) and Xe target (right). The different color intensities and thicknesses of the lines are from light thin to dark thick: 1 eV, 10 eV, 100 eV, 1 keV, 2.5 keV, 5 keV, 7.5 keV, and 10 keV.} 
\label{fig:WIMPSpectrum}
\end{center}
\end{figure*} 

As shown in Fig.~\ref{fig:1neutrinoLine} (left), there is a large change in the WIMP sensitivity corresponding to exposures leading to one neutrino event between WIMP masses of 5 to 10~GeV/c$^2$. This is due to the fact that a Xe-based experiment needs to have a threshold below 4~keV to have sensitivity to WIMPs below $\sim$10~GeV/c$^2$. Below 4~keV the $^8$B and $hep$ neutrinos start to leak into the signal region, and their much larger rate implies a much lower exposure to attain one neutrino event. Conversely, for WIMP masses above $\sim$10~GeV/c$^2$, a Xe-based experiment can achieve better WIMP sensitivity by increasing its threshold above 4~keV to be insensitive to the solar neutrinos and thus has atmospheric neutrinos as its dominant neutrino background. The much lower flux implies a much larger exposure to attain one neutrino event. One can deduce from Fig.~\ref{fig:1neutrinoLine} that aiming at detecting a light WIMP (below 10~GeV/c$^2$) with a cross section below 10$^{-45}$~cm$^2$ or a WIMP heavier than 20~GeV/c$^2$ with a cross section below 10$^{-48}$~cm$^2$ will be very challenging due to the presence of neutrino background, see Sec.~\ref{sec:Discovery}. This new estimation of the neutrino background contamination from background-free exclusion sensitivity limits shown in Fig.~\ref{fig:1neutrinoLine} can also be done for different target nuclei and with energy-dependent detection efficiencies. For lighter targets, the abrupt drop around 6~GeV/c$^2$ will occur at slightly larger masses. This kinematic effect is the same mechanism at work in Fig.~\ref{fig:Targets}, which will be discussed in the next section.
 

\section{WIMP reconstruction of neutrino only data}
\label{sec:Reconstruction}

As the upcoming ton-scale experiments will be sensitive to the neutrino background, it is worth investigating how such a false positive dark matter detection signal could be interpreted in the context of a WIMP only reconstruction of the data. For these purposes, we introduce the WIMP only likelihood function defined as follows \cite{cowan}:
\begin{equation}
\mathscr{L}(m_{\chi},\sigma_{\chi-n}) = \frac{\mu_\chi^{\rm N}}{\rm N!}e^{-\mu_\chi}\prod_{i=1}^{\rm N}f_\chi(E_{r_i}), 
\end{equation}
where $f_{\chi}$ is the unit normalized energy distribution for WIMP-induced nuclear recoils and $\mu_\chi$ is the expected number of WIMP events for a given WIMP mass and WIMP-nucleon cross section ($\sigma_{\chi-n}$)  defined as:
\begin{equation}
\mu_\chi = \int_{E_{\rm th}}^{E_{\rm up}}\frac{dR}{dE_r}dE_r, 
\end{equation}
where $E_{\rm th}$ is the nuclear recoil energy threshold and $E_{\rm up}$ is  the upper bound which is taken to be equal to 100 keV. 

In order to study how a neutrino signal could be interpreted as a potential dark matter signal, we computed the maximum likelihood distribution of 10,000 Monte Carlo pseudo-experiments for which we have set $\sigma_{\chi - n} = 0$~cm$^2$ such that the fake data sets only contain neutrino-induced nuclear recoils. Also, we have varied the total exposure such that the expected number of neutrino events for each threshold energies was about 500 events, which is roughly the number of neutrino events expected for a 0.1 keV threshold Ge experiment with a 1 ton-year exposure.\\

 \begin{figure*}[t]
\begin{center}
\includegraphics[scale=0.44,angle=0]{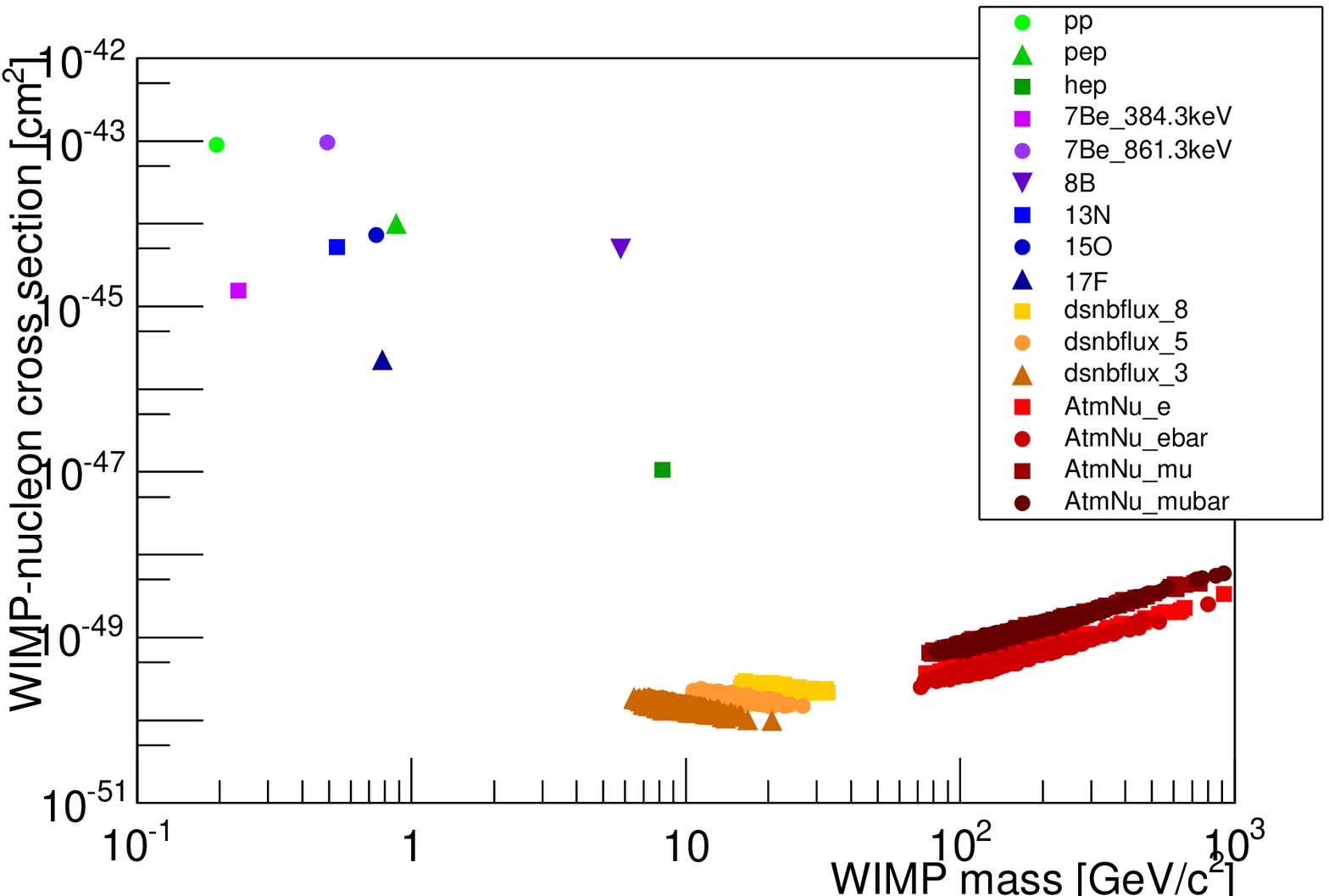}
\includegraphics[scale=0.44,angle=0]{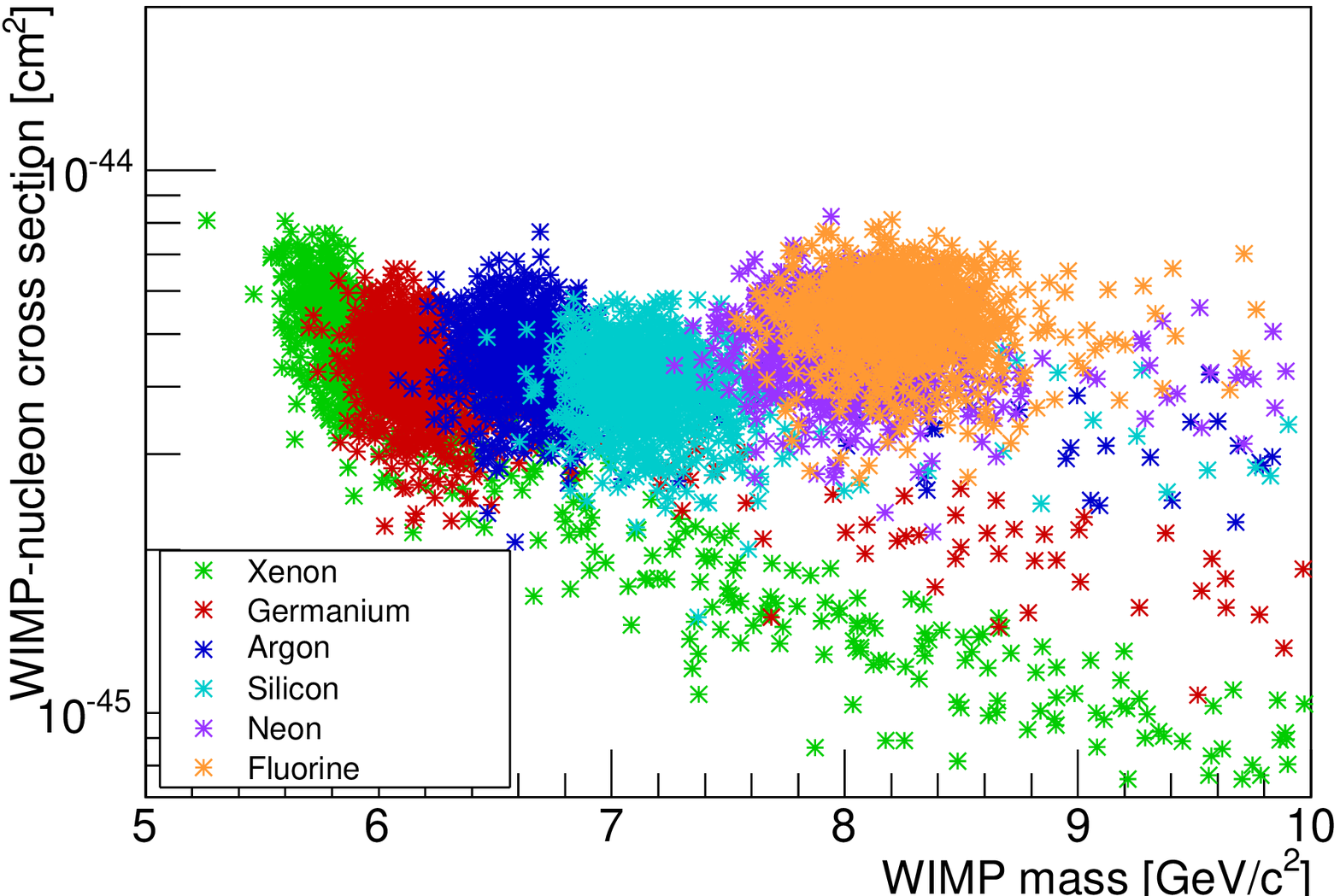}
\caption{Left: Distributions of the maximum likelihood of the CNS background under the WIMP only hypothesis for each neutrino component, considering a Xe target nucleus and no energy threshold. Right: Distributions of the maximum likelihood of the CNS background under the WIMP only hypothesis for six different target nuclei: Xe, Ge, Ar, Si, Ne, F with a common energy threshold of 1 keV. Distributions shown on left and right panels have been computed by adjusting the experiment exposure such that we have a total of about 500 expected neutrino events for each configuration.} 
\label{fig:Targets}
\end{center}
\end{figure*}

The resulting distributions for various energy thresholds are presented in Fig.~\ref{fig:MaxLike} for a Ge target (left panel) and Xe target (right panel) where the different intensities of the coloring correspond to the various energy thresholds considered: 1 eV, 10 eV, 100 eV, 1 keV, 2.5 keV, 5 keV, 7.5 keV, and 10 keV. From the different distributions, one can deduce that for energy thresholds of the order of 1 keV and below, the reconstructed WIMP mass from neutrino background only data should lie within the range of 3 to 30 GeV/c$^2$ in the case of Ge- and Xe-based experiments. The general tendency when increasing the energy threshold is that the reconstructed WIMP mass gets higher and the cross section lower. The first effect  is easily explained by the fact that when the energy threshold increases, the experiment is less sensitive to the lower-energy (but higher flux) neutrinos, and thus the higher-energy neutrinos have a more dominant role, inducing a larger fraction of higher recoil energies which mimics higher WIMP masses. The reduction of the reconstructed cross section comes from the fact that the CNS background is composed of several components that have different end point energies, inducing significant reductions of the event rate when increasing the energy threshold. As a matter of fact, as the reconstructed WIMP mass and cross section drastically depend on the energy threshold, this suggests that the total CNS spectrum is not well fitted by a WIMP only hypothesis on the whole energy range from 1~eV to 100~keV.\\

In order to assess how well the CNS spectrum is fitted by a WIMP only hypothesis, we show in Fig.~\ref{fig:WIMPSpectrum} the total CNS-induced nuclear recoil energy spectrum (blue dashed line) to which is superimposed the mean best fit models for each of the energy threshold configurations where the intensity of the coloring corresponds to the various energy thresholds considered. The WIMP only hypothesis only fits the total CNS background reasonable well for threshold energies above 0.1 keV. Indeed, in the case of a threshold of 1 eV one can see from Figs.~\ref{fig:Rates} and \ref{fig:Number} that the total CNS spectrum is composed of different components that have mainly four different recoil energy end points at roughly 5 eV, 20 eV, 100 eV and 1 keV. As the $pp$ component only dominates by about one order of magnitude, the remaining leading components will have a significant contribution to the recoil energy distribution, resulting in the fact that the WIMP only model does not fit very well the total neutrino spectrum. In the case of a 10 eV energy threshold, the same argument applies even if the resulting total CNS spectrum has only three distinct end points at 20 eV, 100 eV and 1 keV.

These results suggest that the neutrino background could only mimic very well a WIMP detection in the case where the energy threshold is high enough so there is only one very dominant contribution or a smooth superposition of different neutrino components, such as $^8$B and $hep$ neutrinos or atmospheric and diffuse supernova neutrinos. Also, in order to disentangle a neutrino background detection from a true WIMP signal, one could vary the energy threshold of the experiment to get a consistency check of the WIMP hypothesis.\\

As the neutrino background could very well mimic a possible WIMP signal, we could also evaluate to what WIMP model a given neutrino component is equivalent. This is shown in Fig.~\ref{fig:Targets} (left panel) where we present the distributions of maximum likelihood in the WIMP parameter space that are deduced from a given neutrino component. These results have been computed for a Xe target nucleus with no energy threshold. As one can see from this figure, the solar neutrinos tend to be reconstructed at low WIMP masses with high cross sections while the DSNB and atmospheric neutrinos are at much higher WIMP masses and much lower cross sections. From this figure, one can easily deduce that the neutrino background will start becoming important when the experiment will start to reach sensitivities down to $10^{-45}$~cm$^2$ ($10^{-48}$~cm$^2$) for the light (heavy) WIMP range. \\
Figure~\ref{fig:Targets} (right) presents the WIMP reconstructed neutrino backgrounds on the WIMP-nucleon cross section vs.~WIMP mass plane for different target nuclei and a common  energy threshold of 1 keV. With such an energy threshold, the $^8$B and $hep$ neutrinos are the dominant components of the simulated data. For heavier targets such as Xe or Ge, atmospheric and supernova neutrinos have a non-negligible contribution, thus explaining the tails of the distributions at higher WIMP masses. The reconstructed cross section is roughly the same for all targets while the reconstructed WIMP mass is shifted to lower WIMP masses for heavier targets. Interestingly, the fact that the reconstructed WIMP parameters are not strictly identical for each target suggests the possibility to disentangle a neutrino background from a genuine WIMP detection using different target nuclei. However, as the reconstructed parameters are fairly close to each other within statistical uncertainties, one can get the sense that such target complementarity might not be of great help to reduce the impact of neutrino backgrounds on the reach of upcoming ton-scale experiments. Nevertheless, in the case of non-standard WIMP-nucleus interaction such as isospin violating dark matter \cite{Feng:2011vu}, the use of different targets could very efficiently remove degeneracies between a WIMP signal and neutrino backgrounds.


\section{Discovery potential of ton-scale experiments}
\label{sec:Discovery}

As next generation experiments plan to reach the ton scale exposure mass with low thresholds between 0.1 to 2~keV, it is important to assess the discovery potential of such low threshold experiments in the light of neutrino backgrounds.

To provide this assessment we compute the discovery limits for direct detection experiments. Discovery limits were first introduced in Ref.~\cite{Billard:2011zj} and are defined such that if the true WIMP model lies above this limit then a given experiment has a 90\% probability to get at least a 3$\sigma$ WIMP detection. Hence, to derive these limits, it is necessary to compute the detection significance associated with different  WIMP models and for each detector configuration. This can be done  using the standard profile likelihood ratio test statistic~\cite{cowan2} where the likelihood function at a fixed WIMP mass ($m_{\chi}$) is defined as,
\begin{align}
\mathscr{L}(\sigma_{\chi-n},\vec{\phi}) &= \frac{e^{-(\mu_\chi+\sum_{j=1}^{n_{\nu}}\mu^j_\nu)}}{\rm N!}\nonumber \\
& \times \prod_{i=1}^{\rm N}\left[\mu_\chi f_\chi(E_{r_i}) + \sum_{j=1}^{n_{\nu}}\mu^j_\nu f^j_\nu(E_{r_i})\right]\nonumber \\
&\times \prod_{i=1}^{n_{\nu}}\mathscr{L}_i(\phi_i),
\end{align}
where $\mu^j_\nu$ and $f^j_\nu$ are respectively the expected number of neutrino background events and the unit normalized nuclear recoil energy distribution from each neutrino contribution. Finally, $\mathscr{L}_i(\phi_i)$ are the individual likelihood functions related to the flux $\phi_i$ of each neutrino component. These individual likelihood functions are parametrized as gaussian distributions with a standard deviation given by the relative uncertainty of the different neutrino flux normalizations as discussed in Sec.~\ref{sec:NeutrinoFluxes}.

 \begin{figure*}[ht]
\begin{center}
\includegraphics[scale=0.44,angle=0]{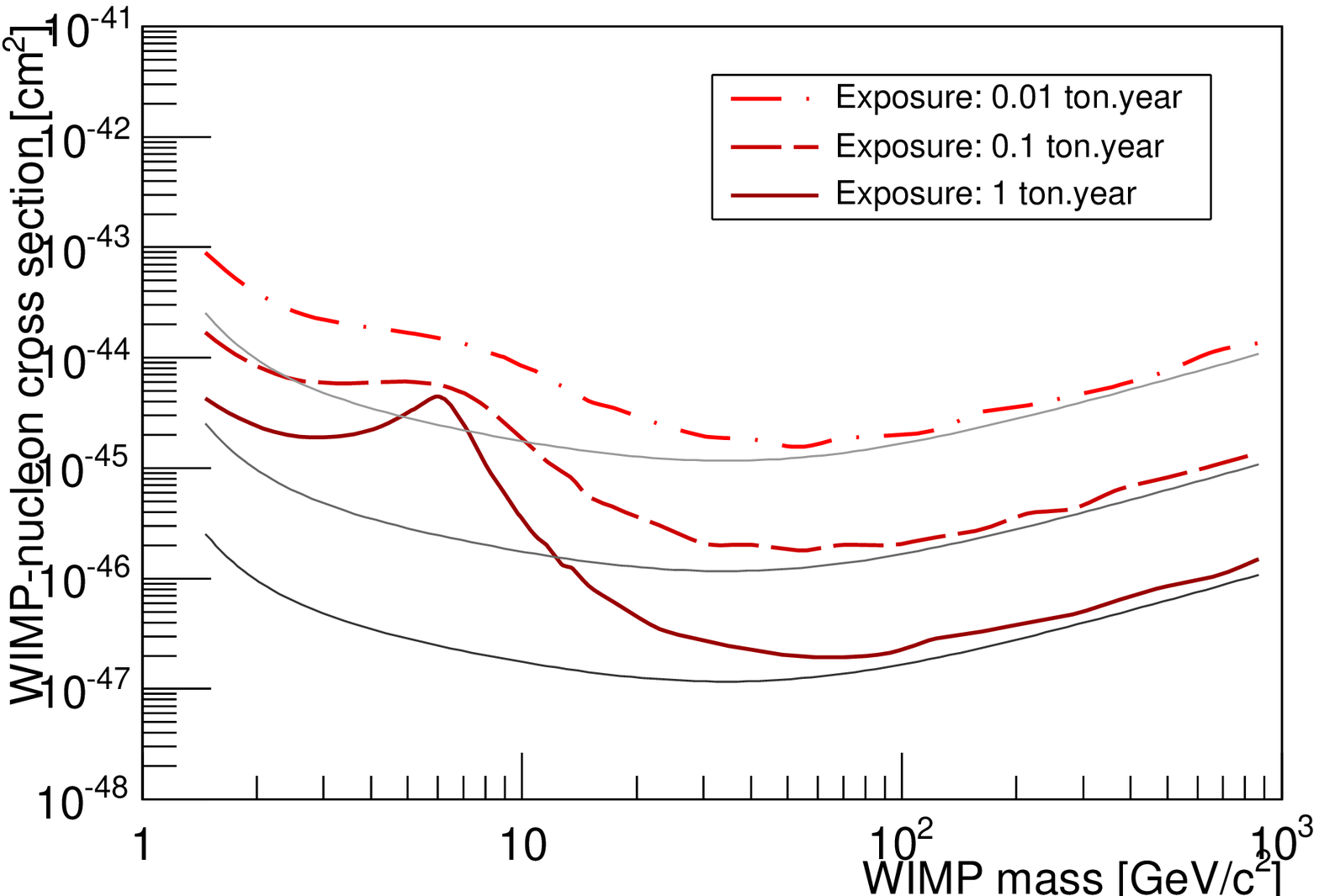}
\includegraphics[scale=0.44,angle=0]{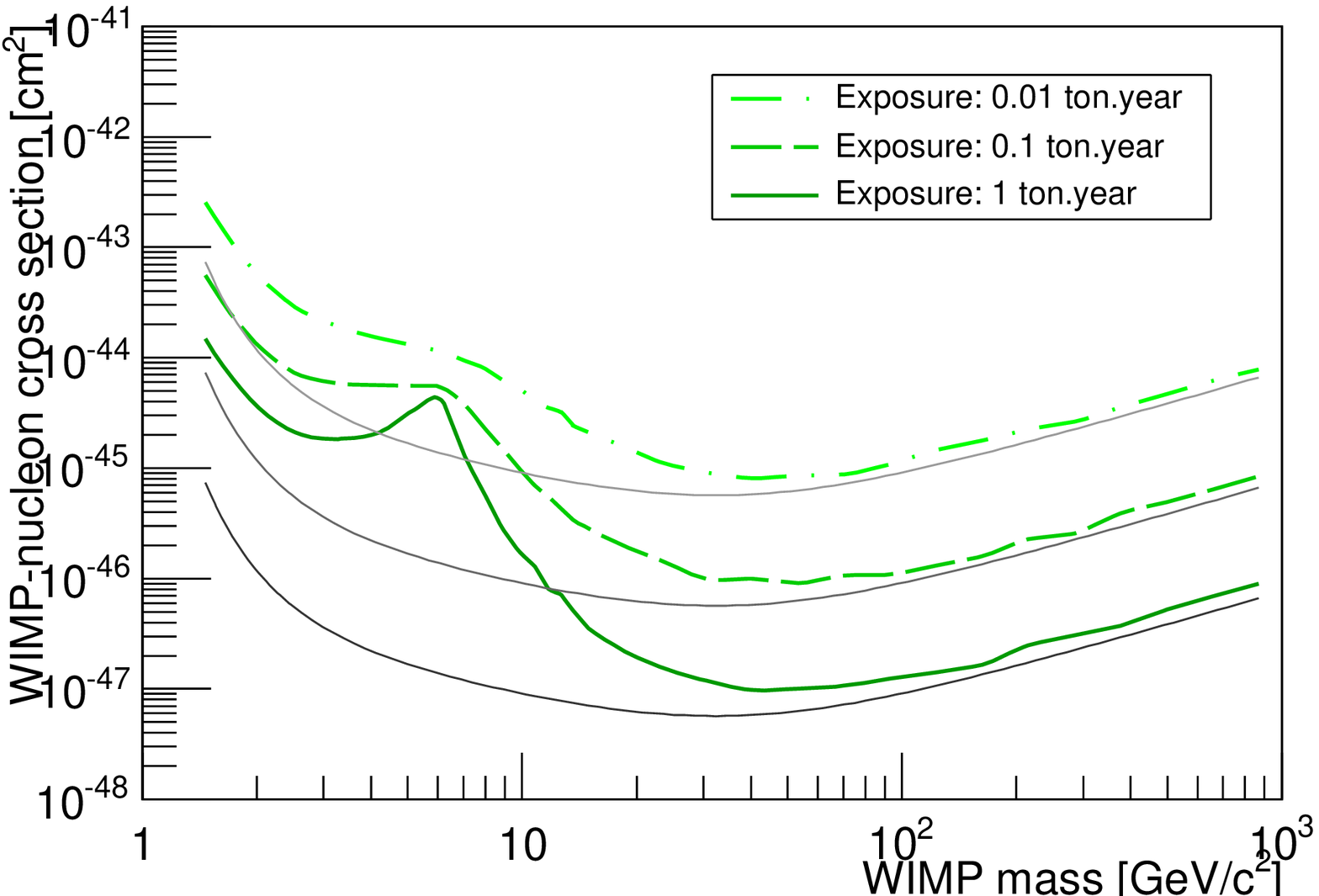}\\
\vspace{0.5cm}
\includegraphics[scale=0.44,angle=0]{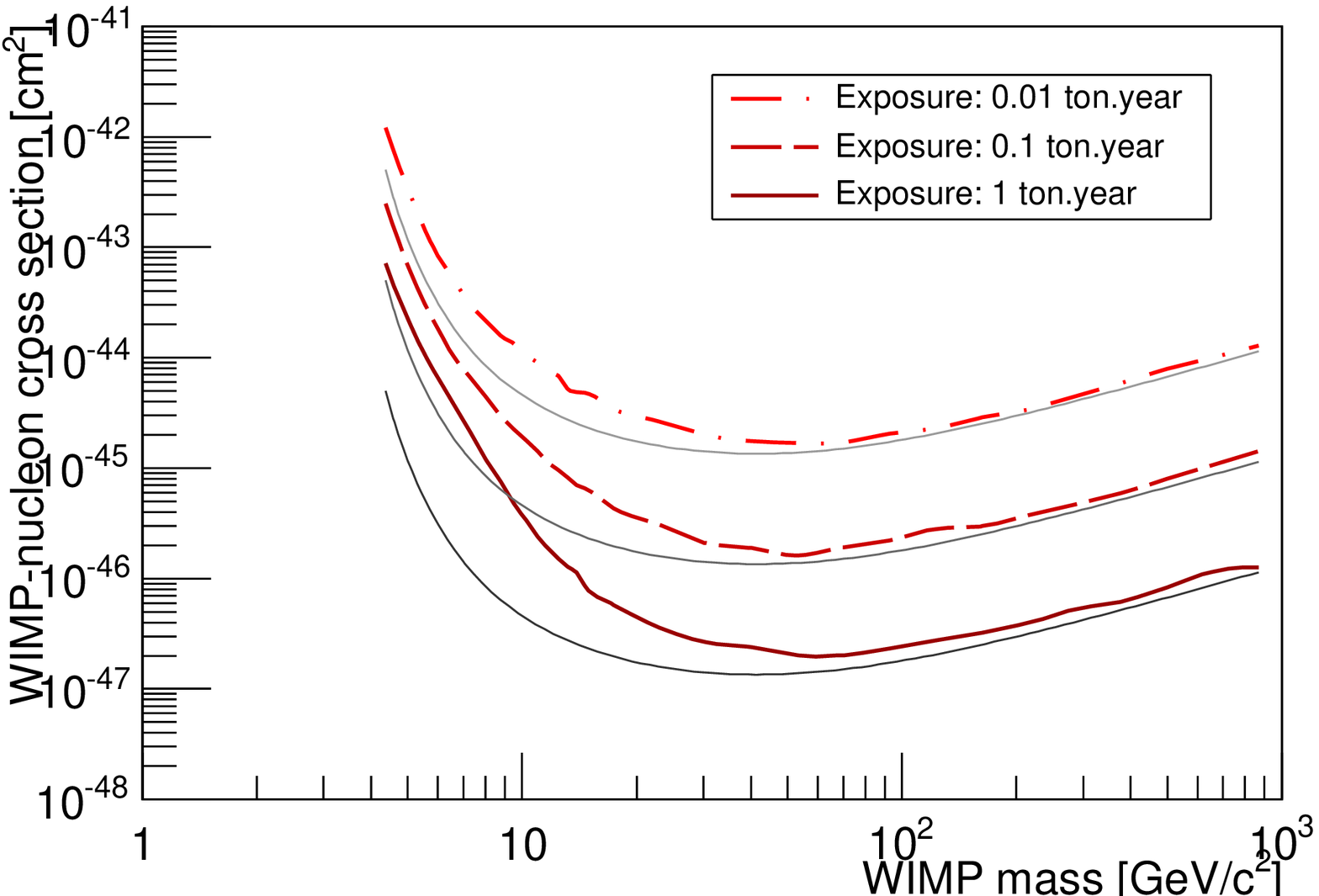}
\includegraphics[scale=0.44,angle=0]{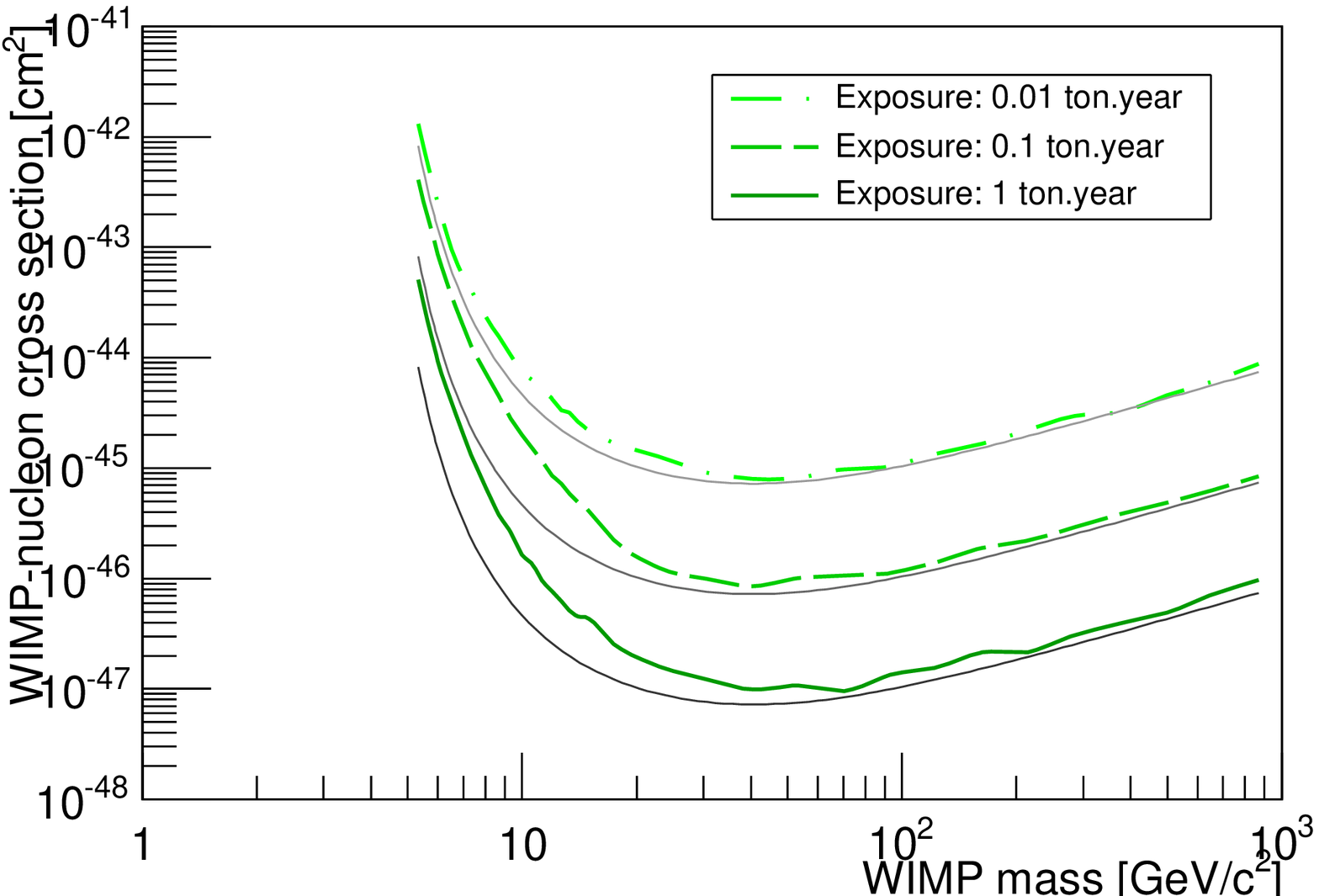}
\caption{Discovery limits for Ge target (left panels) and Xe target (right panels), and for 0.1 keV threshold (top panels) and 2 keV thresholds (bottom panels). The different line styles, solid, dashed, and dotted-dashed respectively correspond to the three exposures 1, 0.1, and 0.01 ton-years. These discovery limits have been computed considering only the two dominant neutrino contributions: $^8$B and $hep$ that both have a 16\% uncertainty on their integrated flux. Also shown are the exclusion sensitivity limits for each of the different exposures with a color intensity scaling with the exposure from light to dark gray. The bumpiness of the discovery limits is due to the finite size of the Monte Carlo samples (500 iterations) inducing a 5\% to 10\% statistical fluctuation over the WIMP mass range.} 
\label{fig:discovery}
\end{center}
\end{figure*} 

The profile likelihood ratio corresponds to a hypothesis test against the null hypothesis $H_0$ (background only) and the alternative $H_1$ which includes both background and signal. Profile likelihood ratio test statistics are designed to  incorporate systematic uncertainties such as the normalization of the neutrino fluxes. As we are interested in the WIMP discovery potential of upcoming experiments, we test the background only hypothesis ($H_0$) on the data and try to reject it using the following likelihood ratio:
\begin{equation}
\lambda(0) = \frac{\mathscr{L}(\sigma_{\chi-n} = 0,\hat{\hat{\vec{\phi}}})}{\mathscr{L}(\hat{\sigma}_{\chi-n},\hat{\vec{\phi}})},
\end{equation}
where $\hat{\hat{\vec{\phi}}}$ denotes the values of ${\vec{\phi}}$ that maximize $\mathscr{L}$ for the specified $\sigma_{\chi-n} = 0$, {\it  i.e.} we are profiling over  ${\vec{\phi}}$ which are considered as nuisance parameters. As discussed in Ref.~\cite{cowan2}, the test statistic $q_0$ is then defined as:
\begin{equation}
q_0 = \left\{
\begin{array}{rrll}
\rm & -2\ln\lambda(0)	&	\ \hat{\sigma}_{\chi-n} > 0 \\
\rm & 0  		& 	\ \hat{\sigma}_{\chi-n} < 0. 
\end{array}\right.
\end{equation}
As one can deduce from such test, a large value of $q_0$ implies a large discrepancy between the two hypotheses, which is in favor of $H_1$ hence a discovery interpretation.
The $p$-value $p_0$ associated to this test is then defined as:
\begin{equation}
p_0 = \int_{q_0^{\rm obs}}^{\infty} f(q_0|H_0)dq_0, 
\end{equation}
where $f(q_0|H_0)$ is the probability distribution function of $q_0$ under the background only hypothesis. Then,  $p_0$ corresponds to the probability to have a discrepancy, between $H_0$ and $H_1$, larger or equal to the observed one $q_0^{\rm obs}$. Following Wilk's theorem, $q_0$ asymptotically follows a $\chi^2$ distribution with one degree of freedom (see Ref.\cite{cowan2} for a more detailed discussion). In such a case, the significance $Z$ in units of sigmas of the detection is simply given by $Z = \sqrt{q^{\rm obs}_0}$.

 \begin{figure}[t]
\begin{center}
\includegraphics[scale=0.44,angle=0]{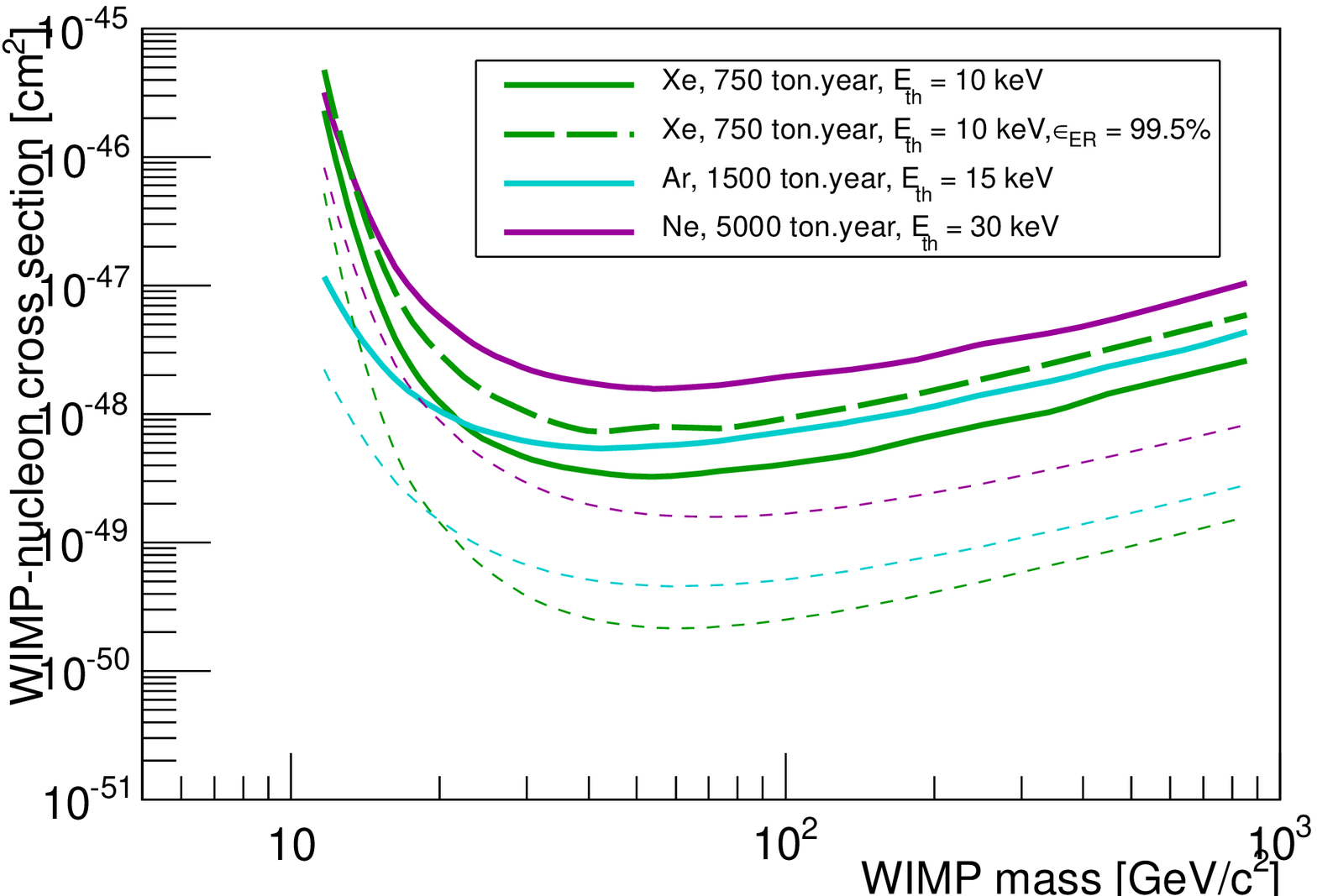}
\caption{Discovery limits (solid lines) for arbitrarily large exposure experiments and for three different target nuclei: Xe (green curves), Ar (cyan curves), and Ne (purple curves). The short dashed lines correspond to the background-free exclusion sensitivity of such experiments. The green long dashed line corresponds to the case where we have considered a finite electron recoil rejection factor of 99.5\% for a XENON-like detector without neutrino magnetic moment enhancement.} 
\label{fig:NeutrinoER}
\end{center}
\end{figure} 

The resulting discovery limits are presented in Fig.~\ref{fig:discovery} in the WIMP-nucleon cross section vs WIMP mass plane for four different experiments: Ge target with a 0.1~keV threshold (top left), Ge target with a 2~keV threshold (bottom left), Xe target with a~0.1 keV threshold (top right), and Xe target with a 2~keV threshold (bottom right). The discovery limits are presented for three different exposures: 10~kg-years, 100~kg-years, and 1 ton-year. Here we have considered only the largely dominant neutrino contributions at these energy thresholds, $^8$B and $hep$. Also shown on Fig.~\ref{fig:discovery} are the background-free exclusion limits for each of the exposures with a color intensity that scales with the exposure from light to dark gray. The  exclusion sensitivity limits help in interpreting the effect of neutrino background on the discovery potential as a function of the WIMP mass.

From Fig.~\ref{fig:discovery}, we can deduce that experiments with a 0.1~keV threshold are significantly affected by the neutrino background. Indeed, we have shown in the previous section that neutrino background could very well mimic a WIMP signal with a mass of $\sim6$~GeV/c$^2$ and a cross section of $\sim5\times 10^{-45}$ cm$^2$. Hence, as  the sensitivity of an experiment gets closer to this point in the WIMP parameter space, the neutrino background starts affecting its discovery potential more significantly. Therefore, near the $\sim6$~GeV/c$^2$ WIMP mass region, one can see that the discovery limit does not evolve linearly with the exposure, but much slower due to the neutrino contamination of the data. It is worth noticing that the energy spectrum induced by the neutrino background and the equivalent $\sim6$~GeV/c$^2$ WIMP are so close to each other that the significance is mainly driven by the theoretical uncertainties on the neutrino flux which are taken to be equal to 16\% for $^8$B and $hep$ neutrinos. One can deduce that smaller uncertainties on the true neutrino flux would allow more accurate background subtraction and thus improve the discovery potential for WIMPs (see Sec.~\ref{sec:map}).

In the case of the 2 keV threshold experiments, the results are completely similar to those previously discussed for high WIMP masses, but they are hardly sensitive to WIMP masses below 10~GeV/c$^2$. Hence the effect of neutrino backgrounds on the discovery limits mainly reduces the sensitivity of the experiment to low WIMP masses. In both energy threshold configurations and for these exposures, the discovery limits are only weakly affected by the CNS backgrounds at high WIMP masses. We have checked that, at these exposures, this holds true even considering the secondary neutrino contributions such as the atmospheric and diffuse supernova neutrinos.

To study the effect of atmospheric and DSNB neutrinos, we explored the case of extremely large exposures to get a sense of the ultimate sensitivity of direct dark matter detection experiments. We considered three different types of target nuclei Xe, Ar, and Ne with energy thresholds of 10, 15, and 30~keV respectively. With such thresholds, only the atmospheric and diffuse supernova neutrinos are relevant as a coherent neutrino scattering background. As shown in Figs.~\ref{fig:MaxLike} and \ref{fig:WIMPSpectrum}, these backgrounds should very well mimic an authentic WIMP signal. Hence, due to their shape similarities with the expected WIMP signal and their relatively large total flux uncertainties, this neutrino background will end up setting a lower limit on the achievable WIMP-nucleon cross section that one could reach with an arbitrarily large dark matter experiment.

This is illustrated in Fig.~\ref{fig:NeutrinoER} where we show the discovery limits (solid lines) of the three different experiments using Xe, Ar and Ne nuclei with a corresponding exposure of 500, 750, and 2500~ton-years respectively. These exposures have been set such that we expect for each experiment about 20 neutrino-induced nuclear recoils. As is shown, the discovery limits are about a little more than one order of magnitude above the corresponding exclusion sensitivity limits (short dashed lines) at high WIMP masses. This is coming from the fact that we have no spectral discrimination and that the discovery limits are almost completely driven by the statistical fluctuations and systematic uncertainties on the total atmospheric and diffuse supernova neutrino fluxes (see Sec.~\ref{sec:map}).  
 
The green long dashed line shown on Fig.~\ref{fig:NeutrinoER} represents the discovery limit when considering the additional neutrino-electron background (without magnetic moment enhancement) with a rejection factor of 99.5\%. It is interesting to note that for the considered Xe exposure, the expected number of neutrino-induced electron recoil is about $\sim$700, which is much greater than the  CNS background. However, as the energy spectrum of the neutrino-electron background is flat over the considered energy range, the spectral discrimination is efficient enough such that the discovery limit is only weakly affected by this additional background contribution by about a factor of 2.

\section{Mapping WIMP Discovery Limit}
\label{sec:map}

 \begin{figure*}[t]
\begin{center}
\includegraphics[scale=0.44,angle=0]{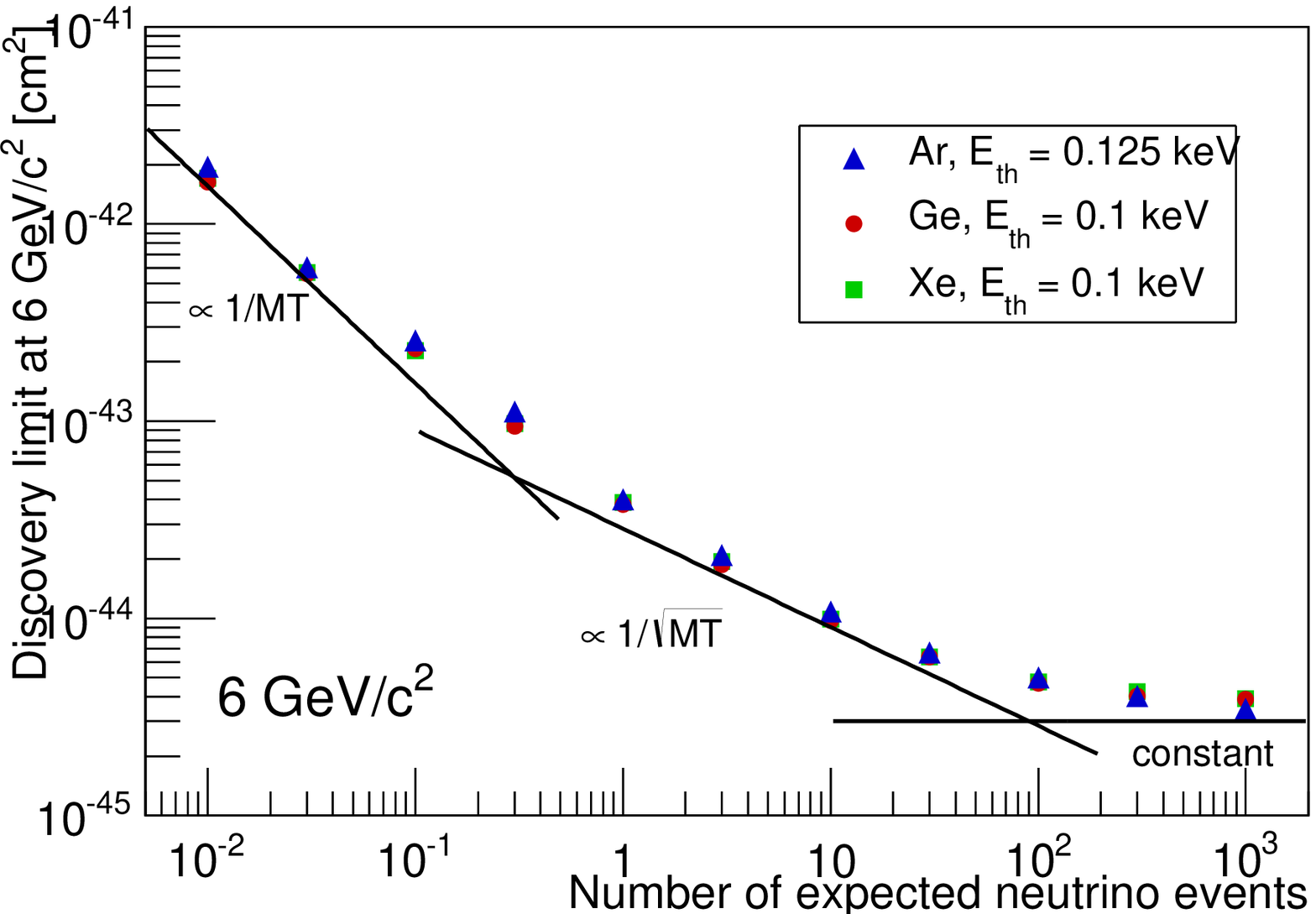}
\includegraphics[scale=0.44,angle=0]{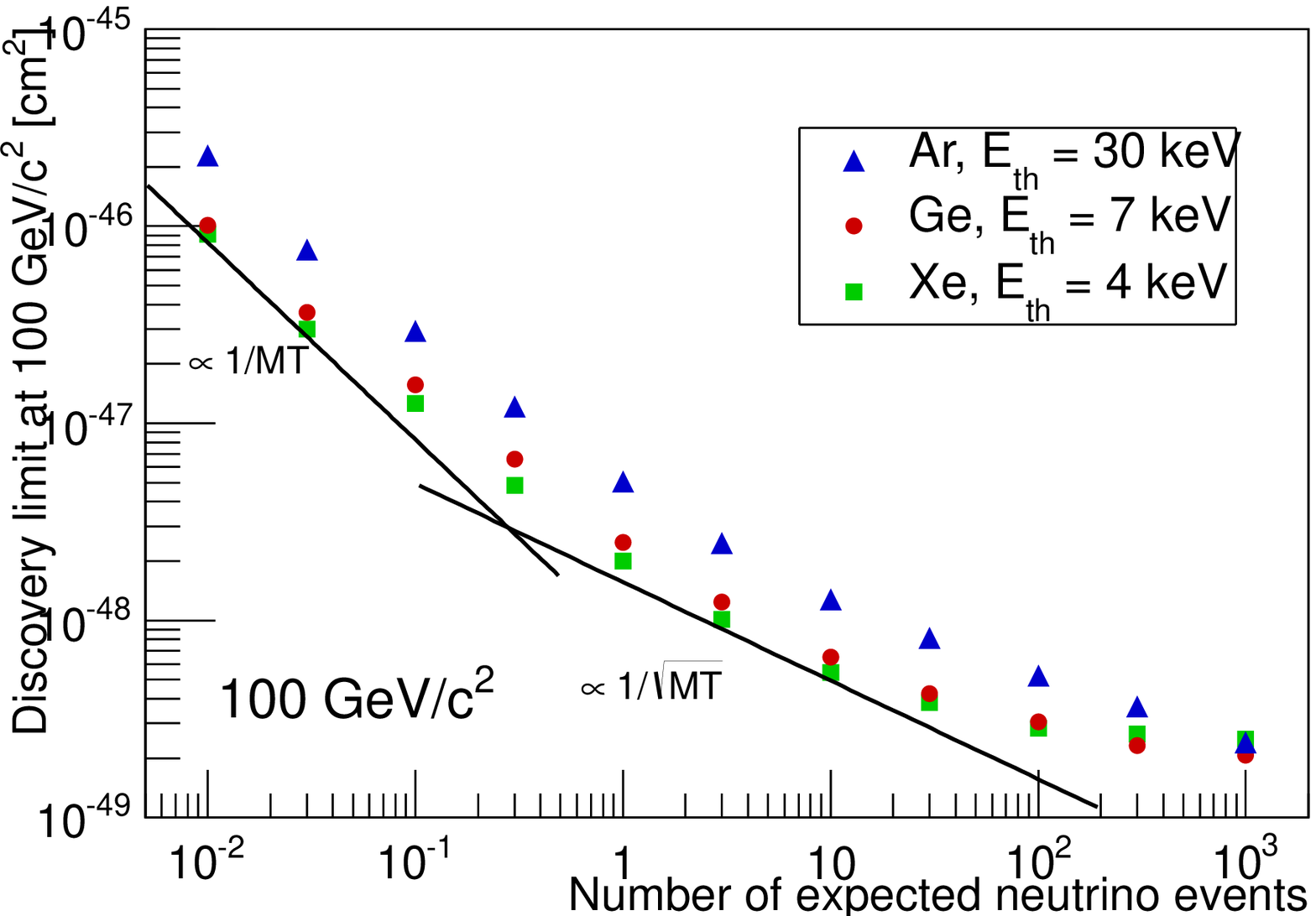}
\caption{Discovery limits for Ar, Ge, and Xe vs exposure. Left: For 6~GeV/c$^2$ WIMPs, the Ge, Xe and Ar exposures required to obtain 100 neutrino events are 240, 130, and 430 kg-years. Right: For 100~GeV/c$^2$ WIMPS, the Ge, Xe, and Ar exposures required to obtain one neutrino event are 32.5, 21.5, and 98 ton-years.} 
\label{fig:transition}
\end{center}
\end{figure*}

As shown in the previous section, when the neutrino background starts to become significant, it can considerably slow down the evolution of the discovery potential with exposure for a given dark matter experiment. Furthermore, we have seen that in the case where a neutrino spectrum and the WIMP spectrum matches very well, as for a WIMP mass of 6~GeV/c$^2$ (see Fig.~\ref{fig:discovery} top panels), the discovery limit starts to saturate. We explore how this discovery potential evolves as a function of exposure in Fig.~\ref{fig:transition}, where we computed the discovery limit at a fixed WIMP mass of 6~GeV/c$^2$ (left panel) and 100~GeV/c$^2$ (right panel) for three different targets and thresholds. As one can see, when the neutrino contribution is negligible, the discovery limit scales as $1/{MT}$. When the neutrino background starts to leak into the signal region, the discovery limit scales as $1/\sqrt{MT}$ if the neutrino and WIMP spectra match sufficiently well such that the discrimination is low. Finally, for even larger neutrino contribution, we can see that the discovery limit slows down even more and eventually becomes constant as a function of exposure. The latter transition is due to the systematic uncertainties on the total neutrino flux. Indeed, in the case of a significant neutrino contribution and a perfect match between the neutrino and WIMP spectra, one can approximate the evolution of the discovery limit as:

\begin{equation}
\sigma_{90\%} \propto \frac{ \sqrt{ N_{\nu} + \xi^2(N_{\nu})^2}}{N_{\nu}} = \sqrt{\frac{1+\xi^{2}N_{\nu}}{N_{\nu}}},
\end{equation}
where $\sigma_{90\%}$ is the 90\% discovery limit, $N_{\nu}$ is the expected number of neutrino events which scales linearly with exposure, and $\xi$ is the systematic error in percentage on the neutrino contribution. When $\xi^{2}N_{\nu} \ll 1$ the cross section scales as $1/\sqrt{N_{\nu}}$ (pure Poisson regime at low number of neutrino events), while when $\xi^{2}N_{\nu} \gg 1$ the cross section becomes constant with increasing exposure (purely dominated by systematics, at high neutrino contamination).

This suggests that for WIMP masses that produce energy spectra that nearly match the CNS background, the systematic uncertainties on the neutrino flux will end up setting a lower limit on the reachable discovery potential of upcoming dark matter experiments. Note that the level of neutrino background for which the discovery limit starts to saturate is directly related to the systematic error $\xi$ on the neutrino flux. Indeed, in the case of a perfect match between WIMP and neutrino spectra, one can easily derive that the exposure at which the transition between Poisson-dominated and systematics-dominated regime occurs is given by: $N_{\nu} = 1/\xi^2$. Therefore, an improvement of a factor of 2 in the systematic uncertainties will postpone the saturation of the discovery limit at an exposure 4 times larger and improve the discovery reach by a factor of 2.

For the 6~GeV/c$^2$ case (Fig.~\ref{fig:transition} left), the exposures required to reach the saturation point around 100 neutrino events are 240 kg-years for Ge, which are exposures accessible to next generation experiments. For the 100~GeV/c$^2$ case, however, the exposure required to get 100 of neutrino background events is 2,150 ton-years. Given these exposure numbers, it is likely that at high masses, in the absence of a WIMP signal at higher cross sections, discovery limits much below $10^{-48}$~cm$^{2}$ will become impractical due to the large exposures required even in the Poisson-dominated regime.

\begin{figure*}[t]
\begin{center}
\includegraphics[height=2.5in]{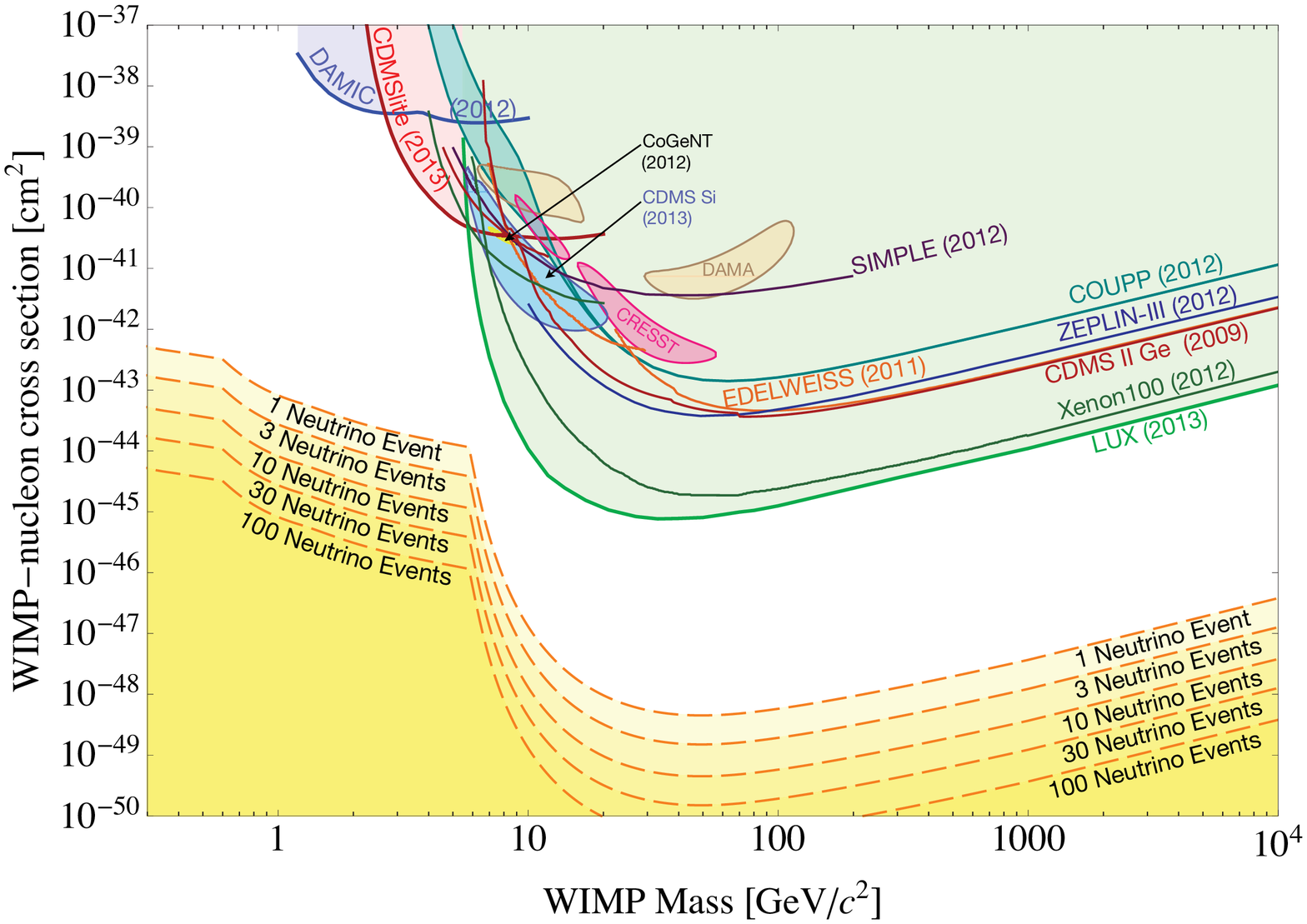}\hspace{-.5in}
\includegraphics[height=2.5in]{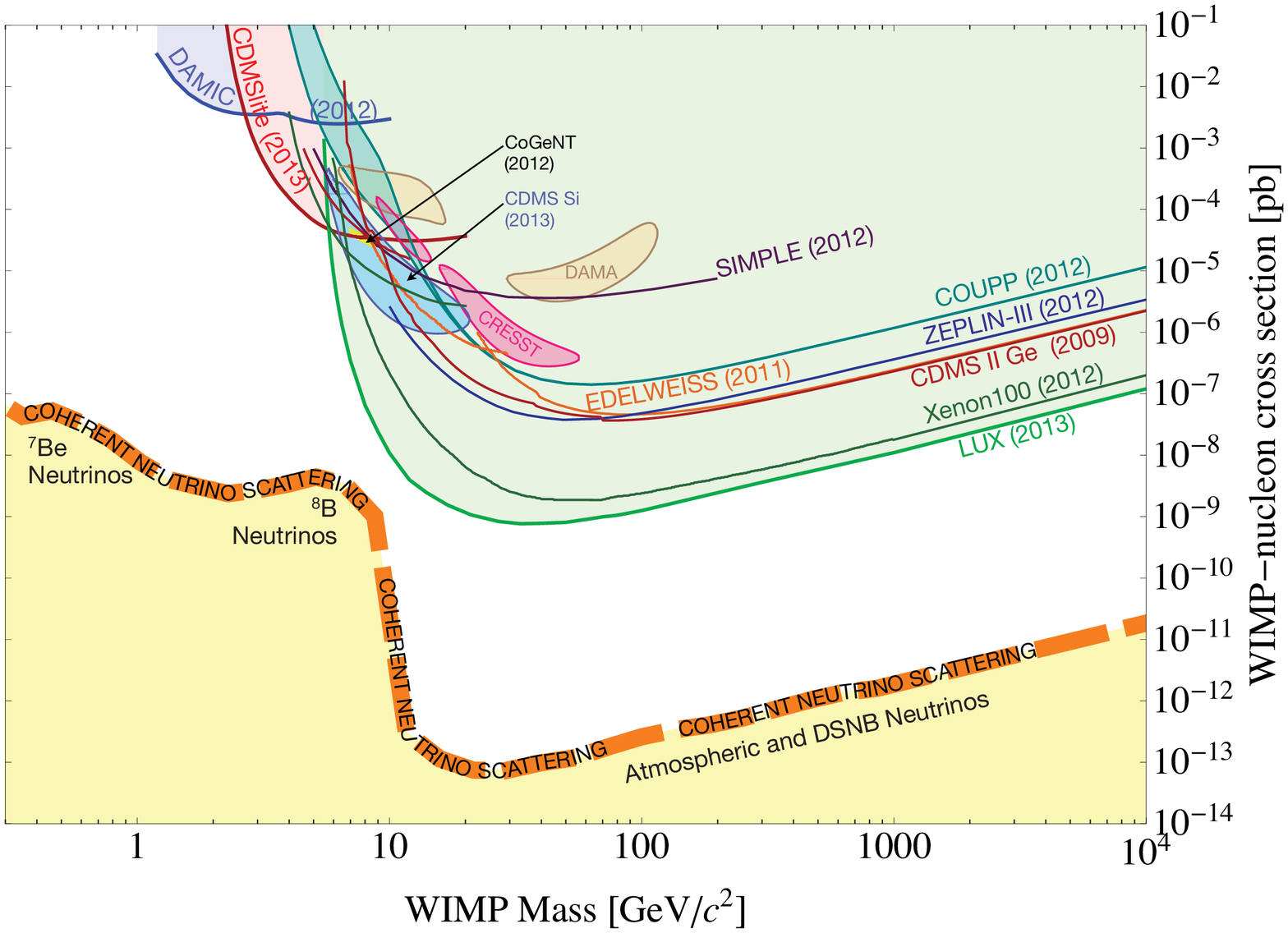}
\caption{Left: Neutrino isoevent contour lines (long dash orange) compared with current limits and regions of interest. The contours delineate regions in the WIMP-nucleon cross section vs WIMP mass plane which for which dark matter experiments will see neutrino events (see Sec.~\ref{sec:nurate}). Right: WIMP discovery limit (thick dashed orange) compared with current limits and regions of interest. The dominant neutrino components for different WIMP mass regions are labeled. Progress beyond this line would require a combination of better knowledge of the neutrino background, annual modulation, and/or directional detection. We show 90\% confidence exclusion limits from DAMIC \cite{Aguilar-Arevalo:2013uua} (light blue), SIMPLE \cite{simple} (purple), COUPP \cite{coupp} (teal), ZEPLIN-III \cite{zeplin} (blue),  EDELWEISS standard \cite{EdelStd} and low-threshold \cite{edel:2012} (orange), CDMS~II~Ge standard \cite{CDMSScience:2010}, low-threshold \cite{Ahmed:low-thresh} and CDMSlite \cite{CDMSLITE} (red), XENON10 S2-only \cite{XENON10S2} and XENON100 \cite{Aprile:2012nq} (dark green) and LUX \cite{Akerib:2013tjd} (light green). The filled regions identify possible signal regions associated with data from CDMS-II Si \cite{Agnese:2013rvf} (light blue, 90\% C.L.), CoGeNT \cite{Aalseth:2012} (yellow, 90\% C.L.), DAMA/LIBRA \cite{Bernabei:2010yi} (tan, 99.7\% C.L.), and CRESST \cite{CRESST:2012}  (pink, 95.45\% C.L.) experiments. The light green shaded region is the parameter space excluded by the LUX Collaboration.} 
\label{fig:WIMPlimit}
\end{center}
\end{figure*}

As a final calculation, we have mapped out the WIMP discovery limit across the 500~MeV/$c^2$ to 10~TeV/$c^2$, shown in Fig.~\ref{fig:WIMPlimit} (right). To cover this large WIMP mass range, we combined the discovery limits of two Xe-based pseudo-expriments with a threshold of 3 eV and 4 keV. To ensure we are well into the systematics limited regime, exposures were increased to obtain 500 neutrino events. This line thus represents a hard lower discovery limit for dark matter experiments. Interestingly, we can denote three distinct features in the discovery limits coming from the combination of $^7$Be and CNO neutrinos, $^8$B and $hep$ neutrinos and atmospheric neutrinos at WIMP masses of 0.5, 6, and above 100~GeV/c$^2$ respectively. Also shown are the current exclusion limits and regions of interest from several experimental groups. If the potential WIMP signals around 10~GeV/$c^2$ are shown not to be from WIMPs, the remaining available parameter space for WIMP discovery is bounded at the top by the LUX Collaboration and at the bottom by the neutrino background.
Progress below this line would require very large exposures, lower systematic errors on the neutrino flux, detection of annual modulation, and/or large directional detection experiments.

\section{Conclusion}
\label{sec:Conclusion} 

We have examined the limitations on the discovery potential of WIMPs in direct detection experiments due to the neutrino backgrounds from the Sun, atmosphere, and supernovae. We have specifically focused on experiments that are only sensitive to energy deposition from WIMPs. 
We have determined the minimum detectable spin-independent cross section as a function of WIMP mass over a wide range of masses from 500~MeV/$c^2$ to 10~TeV/$c^2$ that could lead to a significant dark matter detection. WIMP-nucleon cross sections of $\sim$10$^{-45}$ and $\sim$10$^{-49}$~cm$^2$ are the maximal sensitivity to light and heavy WIMP dark matter respectively that direct detection searches without directional sensitivity could reach, given the uncertainties on the neutrino fluxes. This limit is roughly about 3 to 4 orders of magnitude below the most recent experimental constraints. In the case of light WIMPs (about 6~GeV/c$^2$) next generation experiments might already reach the saturation regime with about 100 neutrino background events. For heavier WIMPs (above 20 GeV/c$^2$) we have shown that progress below 10$^{-48}$ cm$^2$ will be strongly limited by the very large increases in exposure required for decreasing gains in discovery reach. 

\par As a main conclusion of this work, our results show that the cosmic neutrino background poses a hard limit on the discovery potential of future direct detection experiments. However, it is possible to reduce the impact of neutrino backgrounds on direct searches experiments in four ways:

\begin{enumerate}

\item An improvement in the theoretical estimation and experimental determination of the neutrino fluxes. In particular more precise measurements of the different neutrino flux components by future experiments will improve the ultimate discovery limit of dark matter experiments.

\item A utilization of different target nuclei. As we have shown in Fig.~\ref{fig:Targets}, even though utilizing different target nuclei generally does not improve sensitivity as much as an increase in exposure does, it will be important for independent measurements of the neutrino fluxes and the coherent scattering cross section. This is consistent with several recent analyses~\cite{Pato:2010zk,Pato:2012fw}. However, it is certainly likely that if the WIMP couples differently to the proton and neutron, as in the case of isospin-violating dark matter, the utilization of different target nuclei will be even more important.  

\item Measurement of annual modulation. In the case of a 6 GeV/c$^2$ WIMP, next generation experiments could reach sufficiently high statistics to disentangle the WIMP and the neutrino contributions using the 6\% annual modulation rate of dark matter interactions~\cite{Freese:2012xd}. However, in the case of heavier WIMPs, very large and unrealistic exposures would be required to obtain enough events to detect such predicted annual modulation for cross sections around 10$^{-48}$~cm$^2$. Furthermore, the atmospheric neutrino event rate also undergoes annual modulation due to the change in temperature of the atmosphere throughout the year \cite{Tilav:2010}. A dedicated study taking into account systematic uncertainties in the neutrino fluxes and their modulations is required to assess the feasibility of annual modulation discrimination in light of atmospheric neutrino backgrounds.

\item Measurement of the nuclear recoil direction as suggested by upcoming directional detection experiments~\cite{Ahlen:2009ev}. Since the main neutrino background has a solar origin, the directional signal of such events is expected to be drastically different than the WIMP-induced ones~\cite{Billard:2009mf,Billard:2010jh}. This way, a better discrimination between WIMP and neutrino events will  enhance the WIMP detection significance allowing us to get stronger discovery limits.\\
\end{enumerate}

\section*{Acknowledgments}
The authors would like to thank Adam Anderson, Blas Cabrera, Peter Sorensen, Rick Gaitskell, Dan McKinsey, Cristiano Galbiati, and Dan Bauer for useful discussions and for providing insightful comments on the manuscript. This work was funded in part by the National Science Foundation Grant No. NSF-0847342.

\end{document}